\documentclass{aa} 
\usepackage{graphicx}
\usepackage{txfonts}
\begin{document}

\title{The Hubble PanCET program: \\ The near-ultraviolet transmission spectrum of WASP-79b} 
        
\author{
A.~Gressier \inst{1,2,3,10}
\and
A.~Lecavelier des Etangs \inst{2} 
\and 
D.K.~Sing\inst{4,5}
\and
M.~L\'opez-Morales\inst{6}
\and
M.K.~Alam \inst{7}
\and
J.K.~Barstow\inst{8}
\and 
V.~Bourrier \inst{9}
\and 
L.A.~Dos Santos\inst{10}
\and
A.~Garc\'ia Mu\~noz \inst{11}
\and
J.D.~Lothringer\inst{12,5}
\and
N. K.~Nikolov\inst{10}
\and
K.S.~Sotzen \inst{13,4}
\and 
G.W.~Henry\inst{14}
\and
T.~Mikal-Evans\inst{15} 
}

\institute{LATMOS, CNRS, Sorbonne Universit\'e UVSQ, 11 boulevard d’Alembert, F-78280 Guyancourt, France \email{amelie.gressier@latmos.ipsl.fr}\label{inst1}
\and
Institut d'Astrophysique de Paris, CNRS, UMR 7095, Sorbonne Universit\'e, 98 bis bd Arago, 75014 Paris, France \label{inst2}
\and
LESIA, Observatoire de Paris, Universit\'e PSL, CNRS, Sorbonne Universit\'e, Universit\'e de Paris, 5 place Jules Janssen, 92195 Meudon, France\label{inst3}
\and
Department of Earth \& Planetary Sciences, Johns Hopkins University, Baltimore, MD, USA\label{inst4}
\and
Department of Physics \& Astronomy, Johns Hopkins University, Baltimore, MD, USA\label{inst5}
\and
Center for Astrophysics ${\rm \mid}$ Harvard {\rm \&} Smithsonian, 60 Garden St, Cambridge, MA 02138, USA\label{inst6}
\and
Carnegie Earth \& Planets Laboratory, 5241 Broad Branch Road NW, Washington, DC 20015\label{inst7}
\and
School of Physical Sciences, The Open University, Walton Hall, Milton Keynes, MK7 6AA, UK\label{inst8}
\and
Astronomical Observatory, University of Geneva, Chemin Pegasi 51b, CH-1290 Versoix, Switzerland\label{inst9}
\and
Space Telescope Science Institute, 3700 San Martin Drive, Baltimore, MD 21218, USA \email{agressier@stsci.edu}\label{inst10}
\and
Universit\'e Paris-Saclay, Universit\'e Paris Cit\'e, CEA, CNRS, AIM, 91191, Gif-sur-Yvette, France\label{inst11}
\and
Department of Physics, Utah Valley University, Orem, UT 84058, USA\label{inst12}
\and
JHU Applied Physics Laboratory, 11100 Johns Hopkins Road, Laurel, MD 20723 USA\label{inst13}
\and
Center of Excellence in Information Systems, Tennessee State University, Nashville, TN 37209, USA\label{inst14}
\and
Max Planck Institute for Astronomy, Königstuhl 17, D-69117 Heidelberg, Germany\label{inst15}
}

\date{accepted January 31, 2023}
%\date{Received Janvier 53, 2022; accepted March 37, 2022}
   
        %%ABSTRACT
        
        \abstract{
        We present Hubble Space Telescope (HST) transit observations of the Hot-Jupiter WASP-79\,b acquired with the Space Telescope Imaging Spectrograph (STIS) in the near ultraviolet (NUV). Two transit observations, part of the PanCET program, are used to obtain the transmission spectra of the planet between 2280 and 3070 \text{\AA}. 
        We correct for systematic effects in the raw data using the jitter engineering parameters and polynomial modelling to fit the white light curves of the two transits. 
        We observe an increase in the planet-to-star radius ratio at short wavelengths, but no spectrally resolved absorption lines. The difference between the radius ratios at 2400\,\text{\AA} and 3000\,\text{\AA} reaches $0.0191\pm0.0042$ ($\sim$4.5$-\sigma$). 
        Although the NUV transmission spectrum does not show evidence of hydrodynamical escape, 
        the strong atmospheric features are likely due to species at very high altitudes. We performed a 1D simulation of the temperature and composition of WASP-79\,b using Exo-REM.
        The temperature pressure profile crosses condensation curves of radiatively active clouds, particularly MnS, Mg$_2$SiO$_4$, Fe, and Al$_2$O$_3$. Still, none of these species produces the level of observed absorption at short wavelengths and can explain the observed increase in the planet's radius.
        WASP-79 b's transit depth reaches 23 scale height, making it one of the largest spectral features observed in an exoplanet at this temperature ($\sim$1700\,K). 
        The comparison of WASP-79 b's transmission spectrum with three warmer hot Jupiters shows a similar level of absorption to WASP-178\,b and WASP-121\,b between 0.2 and 0.3$\mu$m, while HAT-P-41 b's spectrum is flat. The features could be explained by SiO absorption.  
        }


   \maketitle
%
%-------------------------------------------------------------------

\section{Introduction} \label{sec:0intro}
Exoplanet atmospheric characterisation has been mainly carried out using observations of close-in transiting planets. Transit spectroscopy in the visible or in the infrared enabled the detection of atomic and molecular species \citep{Deming_2013, Kreidberg_2014b,  Sing_2016, Tsiaras_2018, Madhusudhan_2019}, as well as clouds and hazes \citep{Pont_2008, Lecavelier_2008a, Bean_2010, Kreidberg_2014a} in the lower parts of the atmosphere. Absorption spectroscopy in the ultraviolet (UV) gives access to the upper part of the atmosphere up to the exosphere (10$^{-6}$-10$^{-7}$ bar) and enables the detection of atomic and ionic species \citep{Redfield_2008,  Fossati_2010, Wyttenbach_2015, Arcangeli_2018, Spake_2018}. Exoplanets orbiting close to their star receive stellar X-ray and extreme UV radiations leading to heating at the base of the thermosphere and to the hydrodynamical expansion of the upper layers \citep{Watson_1981, Lecavelier_2004, Murray_clay_2009, Erkaev_2016, Salz_2016}. Highly irradiated hot-Jupiter and Netpune-mass planets are sensitive to hydrodynamic outflows and atmospheric mass escape \citep{Bourrier_2013b, Bourrier_2018, Ehrenreich_2015, Owen_2019}. 

Lyman-$\rm \alpha$ measurements in the far-UV (FUV) revealed an extended hydrogen atmosphere around HD 209458b \citep{Vidal_Madjar_2003}, HD 189733b \citep{Lecavelier_des_Etangs_2010, Lecavelier_des_Etangs_2012, Bourrier_2013a}, GJ 436b \citep{Kulow_2014, Ehrenreich_2015, Lavie_2017, dos_Santos_2019}, GJ 3470b \citep{Bourrier_2018}, and HAT-P-11\,b \citep{Allart_2018, Ben_Jaffel_2022, Dos_Santos_2022} producing a much larger transit depth (10\% to 50\%) than the transit of the planet at optical and infrared (IR) wavelengths (0.1-1$\%$). The atmospheric outflow carries heavier elements from the 0.1–1 bar atmospheric level up to the thermosphere \citep{Garcia_Munoz_2007}. The velocities of metal species range between a few hundred metres per second to several kilometres per second at the base of the exosphere and escape the planet's gravitational pull. Those heavy species are detectable in the near-UV (NUV) absorption spectroscopy. During the transit, they cover 2\% to 10\% of the stellar disc. Atomic oxygen, magnesium, and iron, as well ionised carbon, magnesium, and iron have been detected on HD 209458b \citep{Vidal_Madjar_2004, Linsky_2010}, HD 189733b \citep{Redfield_2008, Ben_Jaffel_2013, Wyttenbach_2015}, and WASP-121b \citep{Sing_2019}. Helium was detected in the atmosphere of WASP-107\,b %with more than 4.5 $\sigma$ 
using Hubble Space Telescope Grism102 (HST G102) measurements in the near-infrared (NIR) \citep{Spake_2018}. \citet{Nikolov_2022} recently observed the complete pressure-broadened profile of the
sodium absorption feature in the cloud-free atmosphere of the hot Saturn, WASP-96 b, with the Very Large Telescope. They were able to measure a precise absolute sodium abundance for this planet.

The Space Telescope Imaging Spectrograph (STIS) NUV observations presented here aim to help us understand the connection between the formation of heavy elements through haze dissociation \citep{Parmentier_2015, Parmentier_2016} and the upper expanding atmosphere \citep{Bourrier_2013a, Bourrier_2014a, Bourrier_2014b} of WASP-79\,b and hot Jupiters, in general. Deep clouds cannot be observed directly but can be detected through 
Rayleigh scattering \citep{Lecavelier_2008a, Lecavelier_2008b, de_Mooij_2013a, Dragomir_2015}. Moreover, observing heavy elements at high altitudes can provide information on physical processes lower in the atmosphere. 

\begin{table}[htpb]
        \caption{System and transit parameters used in the analysis. }         
        \label{table:1parameters}     
        \centering                                     
        \begin{tabular}{l | c}          
                \hline\hline                       
                Parameters & Values\\    
                \hline 
                Spectral Type &  F5 \\
                R$_{*}$ [R$_{\odot}$]  & $1.51^{+0.04}_{-0.03}$  \\ 
                M$_{*}$ [M$_{\odot}$] &  $1.39\pm0.06$ \\ 
                $[$Fe/H$]_{*}$ &  $0.03\pm 0.10$ \\
                T$\rm eff*$ [K] &$6600\pm100$ \\     
                log$_{10}$ $g_{*}$ [cgs] & $4.20 \pm 0.15$ \\

                \hline 
                R$_{\rm P}$ [R$_{\rm Jup}$] &$1.53\pm 0.04$  \\
                M$_{\rm P}$ [M$_{\rm Jup}$] &  $0.85\pm 0.08$ \\
                a [AU] & $0.0519\pm0.0008$ \\
                T$_{\rm eq}$[K]\tablefootmark{a} &$1716.2^{+25.8}_{-24.4}$ \\
                \hline
                P [days] & $3.662392\pm0.000004$        \\
                i [deg] & $86.1\pm 0.2$ \\
                e [deg] & 0.0 \\
                $\omega$ [deg] & 90.0 \\
            a/R$_\star$  & $7.407\pm0.109$ \\
            R$_{\rm P}$/R$_\star$ & $0.10440\pm0.00048$ \\
            %Transit depth &  $0.01148\pm 0.00051$\\
                \hline  
                Reference &\citet{Brown_2017}  \\ 
        \end{tabular}
        \tablefoot{ \tablefoottext{a}{Assuming full heat redistribution and albedo equal to zero.}}
\end{table}

WASP-79b was discovered by \citet{Smalley_2012} with the Wide Angle Space Telescope (WASP-South) and the Transiting Planets and Planetesimals Small Telescope (TRAPPIST). Using a main-sequence mass-radius constraint on the Markov chain Monte Carlo (MCMC) process, they found a mass of 0.90$\pm$0.08M$_{\rm J}$ and a radius of 1.70$\pm$0.11R$_{\rm J}$. The radius was found to be larger, 2.09$\pm$0.14R$_{\rm J}$ while using a non-main-sequence constraint. \citet{Brown_2017} refined the parameters and found a planetary radius of 1.53$\pm$0.04R$_{\rm J}$ and a mass of 0.85$\pm$0.08M$_{\rm J}$ 
(Table~\ref{table:1parameters}). 
The large radius and the mass close to one Jupiter mass yield a low density of $\rho\sim$0.31\,g\,cm$^{-3}$, suggesting an inflated atmosphere. 
The planet orbits its F-type star in 3.7 days and has a high equilibrium temperature of $\sim$1700 K. 
\citet{Addison_2013} showed that the Rossiter-McLaughlin signal of the planet suggests a nearly polar orbit. 
\citet{Sotzen_2020} analysed HST Wide Field Camera 3 (WFC3) and Magellan Low Dispersion Survey Spectrograph (LDSS-3C) data along with Spitzer data and reported the 0.6 to 4.5$\rm \mu$m transmission spectrum of WASP-79 b. 
They found evidence of water vapour and presented a retrieval analysis that favours the presence of FeH and H$^-$ in the atmosphere. 
Another independent study on HST WFC3 Grism 141 (G141) data confirmed the water detection and the possible presence of iron hydride \citep{Skaf_2020}. More recently, a full 0.3 to 5.0\,$\mu$m spectrum was published by \citet{rathcke2021hst} with an analysis of HST PanCET data between 0.3 and 1.0\,$\mu$m. In this work, the transmission spectrum blueward of 1.0\,$\mu$m decreases towards shorter wavelengths with no evidence of hazes or Rayleigh scattering in the planet's atmosphere. On the other hand, they confirmed the water detection with more than 4-$\sigma$ confidence and displayed a moderate detection of H$^{-}$ with 3.3$\sigma$ significance. 
Finally, they detected the effect of unocculted stellar facul\ae\ on the observed spectrum of the planet's atmosphere.

\section{Observations}
\label{sec:1observations}

Observations of WASP-79b used in the present analysis are part of the Panchromatic Exoplanet Treasury (PanCET) program (HST GO Proposal \#14767, PI D.K.~Sing and M.~Lopez-Morales). 
We aim to complete the WASP-79\,b spectrum proposed in \citet{Sotzen_2020} and \citet{rathcke2021hst} by adding UV measurements between 0.2 and 0.3\,$\mu$m. We analysed two transits with the Hubble Space Telescope Imaging Spectrograph (HST STIS) instrument. The observations were conducted with the NUV-Multi-Anode Microchannel Array (NUV-MAMA) detector using the E230M echelle grating with a 0.2"x0.2" aperture. E230M has a resolving power of about 30\,000 and covers the wavelength range of 2280 to 3070\AA. The spectra have an average dispersion of 0.049\AA\ per pixel, about two pixels per resolution element. We observed one transit in each of the two visits of WASP-79b, numbered \#65 and \#66 of the PanCET program. Each observation consisted of five consecutive orbits covering the full transit with a significant baseline. These two observations were taken on January 12, 2018, and March 11, 2018, for the first and second visits, respectively.

The data were acquired in time-tag mode. We extracted a sequence of 350-second-long sub-exposures from these using \verb+calstis+ version~3.4. 
From the shorter observation of the first orbit of each visit, we obtained six~sub-exposures, and from the long observation of the second to the fifth orbit, we obtained eight~sub-exposures per orbit. 
This yields a total of 38~sub-exposures per visit.

\section{Data analysis} \label{sec:2data analysis}
%\subsection{Data reduction}

\subsection{Systematics correction}
\label{sec:Systematics correction}

Spectrophotometric light curves taken with STIS aboard the Hubble Space Telescope (HST) are highly affected by instrument-related systematics. For instance, the thermal 'breathing' effect periodically modifies the point-spread function (PSF). Systematic effects are correlated to instrumental parameters or external factors that vary with time during the observations \citep{Sing_2019}. Before the work of \cite{Sing_2019}, only the thermal breathing effect characterised by the correlation with the HST orbital phase was taken into account. However, corrections limited to only these systematic effects are not enough for data that present a high level of correlated noise, especially for the first HST orbit that presents different systematic effects. 
This led to removing all the data of this first orbit from the analysed data set. 

Given that the first observation of WASP-79 shows a high level of systemic noise (Fig.~\ref{fig:1wlc}),  we implemented a method to analyse and correct systematic effects in HST STIS E230M data using all the information available. Similarly to \citet{Sing_2019}, we decided to include the eighteen measured parameters (see Table~\ref{table:Jitter_details}) that describe HST's Pointing Control System performance during the observation run in the systematic effects
correction model. Those parameters are created by the Engineering Data Processing System (EDPS).

For computational efficiency, we normalised each measurement by subtracting the mean and dividing it by the standard deviation. We did not take into account the HST's latitude and longitude in the detrending model to avoid degeneracy with the HST's phase, with which they are strongly correlated. We decided to keep the first HST orbit, but we discarded the first exposure of each orbit that systematically presents a lower flux level.

    \begin{table}[htb!]
        \caption{Details of the jitter engineering parameters used in the present analysis.
        The parameters from {\tt v2\_dom} to {\tt si\_v3\_p2p} are tabulated in arcseconds. 
        The parameters from {\tt ra} to {\tt los\_zenith} are tabulated in degrees. 
        The magnetic field strengths tabulated in the last three rows are tabulated in Gaussian measures.}
        \label{table:Jitter_details}     
        \centering                                     
        \begin{tabular}{l| l }          
                \hline\hline                       
                Parameters & Meaning  \\    
                \hline 
                \text{\tt v2\_dom} & Dominant guide star V2 coordinate  \\
                \text{\tt v3\_dom} & Dominant guide star V3 coordinate \\ 
                \text{\tt v2\_roll} & Roll FGS V2 coordinate \\ 
                \text{\tt v3\_roll} & Roll FGS V3 coordinate  \\ 
                \text{\tt si\_v2\_avg}  & Mean of jitter in V2 over 3 seconds\\     
                \text{\tt si\_v2\_rms} & Mean of jitter in V2 over 3 seconds\\
                \text{\tt si\_v2\_p2p} & Peak-to-peak jitter in V2 over 3 seconds\\
                \text{\tt si\_v3\_avg}  & Mean of jitter in V3 over 3 seconds\\     
                \text{\tt si\_v3\_rms} & Mean of jitter in V3 over 3 seconds\\
                \text{\tt si\_v3\_p2p} & Peak-to-peak jitter in V3 over 3 seconds\\
                \text{\tt ra} & Right Ascension of aperture reference \\
                \text{\tt dec} & Declination of aperture reference  \\
                \text{\tt roll} &  Position angle between north and +V3 axis \\
            \text{\tt limbang} &  Angle between V1 axis and Earth limb\\
            \text{\tt los\_zenith}  &  Angle between HST zenith and target\\
            \text{\tt mag\_v1}  & Magnetic field along V1  \\
            \text{\tt mag\_v2} & Magnetic field along V2  \\
            \text{\tt mag\_v3} & Magnetic field along V3  \\
\hline
\end{tabular}
\end{table}     

    \begin{table}[htb!]
        \caption{List of the jitter engineering parameters and values of polynomial coefficients included in the white-light curves correction model for the two visits on WASP-79 b. In this case, all polynomials are of degree 1 (see Sect.~\ref{subsec:311wlc}). }  
        \label{table:A1jit}     
        \centering                                     
        \begin{tabular}{l| l | l}          
                \hline\hline                       
                Parameters & Visit \#65\tablefootmark{a} & Visit \#66  \\    
                \hline 
                \text{\tt v2\_dom} & - & -\\
                \text{\tt v3\_dom} & - & - \\ 
                \text{\tt v2\_roll} & - & -\\ 
                \text{\tt v3\_roll}  & - & -\\
                \text{\tt si\_v2\_avg}  & - & -\\     
                \text{\tt si\_v2\_rms} & 5.429995 & -\\
                \text{\tt si\_v2\_p2p} & - & -\\
                \text{\tt si\_v3\_avg}  & - & 0.163022\\     
                \text{\tt si\_v3\_rms} & - & -\\
                \text{\tt si\_v3\_p2p} & - & -\\
                \text{\tt ra} & - & - \\
                \text{\tt dec}  & - & -\\
                \text{\tt roll} & -267.9753 & -\\
            \text{\tt limbang} & - & -\\
            \text{\tt los\_zenith}  & 0.0015797& -\\
            \text{\tt mag\_v1}  & -0.3187546 & -\\
            \text{\tt mag\_v2} & - & -\\
            \text{\tt mag\_v3} & - & -\\
\hline
\end{tabular}
\tablefoot{ \tablefoottext{a}{The correction model of the visit \#65 of the PanCET program also includes a polynomial of degree 8 of the HST orbital phase correlated to the large ramp effect (see Fig.~\ref{fig:1wlc}). The coefficients are the following in increasing degree order: -0.5844569, -1.05257, 17.27887, -20.33507, -37.87583, 251.4479, 7.275825, -1671.329}. We also corrected the linear effect correlated to the planetary phase. The coefficients are 0.12850 and 0.006832 for visit \#65 and visit \#66, respectively.  }
\end{table}     

\begin{table*}[htpb]
        \caption{Limb-darkening coefficients for STIS E230M.}             
        \label{table:2limb}     
%       \centering                                     
        \begin{tabular}{c c | c c c c }          
                \hline\hline 
                 \multicolumn{2}{c|}{Bands} & \multicolumn{4}{c}{Limb darkening coefficient} \\
                 $\lambda_C $(\text{\AA})& $\Delta\lambda $(\text{\AA}) & c1 & c2 & c3 & c4 \\    
                \hline 
        2673 & 799 & 0.44589384 &-0.22547123 &  1.14667951 & -0.42635197\\
        2387 & 236 & 0.41620878 & -0.5781367 & 1.29633765 & -0.1709417\\
        2600 & 200 & 0.53356331 & -0.67401008 & 1.80463859 & -0.70370411\\
        2800 & 200 & 0.46217923 & -0.23965174 & 1.08129327 & -0.3632111\\
        2986 & 172 & 0.39608196 &  0.15363096 & 0.79375784 & -0.42174585\\
        \end{tabular}
\end{table*}

\subsubsection{Detrending parameter selection} \label{subsec:21selection}
The EDPS file contains 18 parameters (Table~\ref{table:Jitter_details}), to which we added the 96 minutes of HST's orbital phase ($\phi_{HST}$). Each parameter describes the configuration of the spacecraft, which varies with time and potentially impacts the photometric measurements. 
To determine whether it is necessary to take a parameter into account in the detrending model, we performed a first fit of the white light curve by testing each parameter independently. For that, we modelled the correlation of the residuals as a function of the parameter value using a first-degree polynomial. The quality of the fit to the light curve was then quantified using the corrected Akaike statistical criterion, $AICc$, which is calculated as a function of k (the number of free variables), n (the total number of exposures), and $\chi^{2}$ the Chi-squared:
\begin{align}
AICc = AIC + \frac{2k^{2}+2k}{n-k-1} ,\\
AIC = \chi^{2}+2\times k.
\end{align}
The effect of a parameter whose correction produces AICc improvement 
(compared to an initial uncorrected adjustment) is integrated into the model. 
We note that we decided to use the corrected AIC rather than the classical AIC or BIC criteria 
because of the small size of the statistical sample, which here is only 33~points (38~sub-exposures minus five~first sub-exposures of each HST orbit).

\subsubsection{Polynomial degree selection}\label{subsec:22polynomial}
Once we identified the parameters to be taken into account to improve the fit to the light curve, we modelled the correction factor as a function of the parameter value using a polynomial function
(Eqs.~\ref{eq:3_1} and~\ref{eq:3_2}) whose degree is chosen to obtain the best fit, that is, the fit to the light curve that produces the lowest AICc value. We obtain a systematic effect correction model composed of a product of polynomials. The parameter values for a given visit are x$_i$ for i ranging from 1 to 19. The number of parameters used in the model depends on the visit.
For each jitter parameter, we increased the polynomial degree $n$ until the AICc 
value no longer decreases from its previous value. 
We also systematically test the $n+1$ degree polynomial if $n-1$ has improved AICc, but $n$ does not. 
At this step, the correction of the tested parameter is incorporated into the light curve fit. The other parameters are tested again (see Section \ref{subsec:21selection}) and include the correction of the identified effects. 
The parameters and polynomial degree selection procedure are carried out iteratively until the systematics correction no longer improves the AICc criterion: 
\begin{align}
 \label{eq:3_1}
 S(X)=\prod_{i=1}^N S_i(x_{i}) ,\\
\label{eq:3_2}
S_i(x_{i}) = 1+a_{1}\times x_{i}+a_{2}\times x^{2}_{i}+...+a_{n}\times x^{n}_{i.}
\end{align}
\label{subsec:23wlcfitting}

\subsection{White light curve fitting} 

We integrated the flux over the entire wavelength range to obtain
a white light curve, with 33~measurements for each of the two observed transits. 
We normalised each light curve with respect to the average flux over a visit, and we modelled the transit
using the \verb+batman+ python package (Basic Transit Model Calculation in Python; \cite{Kreidberg_2015}). 
We held the inclination, semi-major-axis-to-star radius ratio, and limb-darkening coefficients fixed to the value in Tables~\ref{table:1parameters} and~\ref{table:2limb}. 
We modelled the flux measurement over time $f(t)$ as a combination of
the theoretical transit model $T(t,\theta)$, where $\theta$ is the set of the transit parameters, 
the total baseline flux from the host star F$_0$ and the systematic error correction model S(X), derived as described in Sect.~\ref{sec:Systematics correction}:
\begin{equation}
    f(t) = T(t,\theta)\times F_{0}\times S(X)
.\end{equation}
We include a linear baseline time trend in S(X) to correct flux variations over time. The number of free parameters depends on the corrections and S(X) terms. In the search for the best fit, the planet-to-star radius ratio, the polynomial coefficients, and the baseline flux were left free.
We used \verb+batman+'s convention and normalised the time with respect to the centre of the transit. We did not fit the mid-transit time in our analysis 
%of WASP-121\,b and WASP-79\,b light curves 
as that did not improve the fitting results. 
However, our code is flexible, and this parameter can be easily added as a free parameter if necessary.
The best-fit parameters are determined using Levenberg-Marquardt's least-squares method (L-M) \citep{Markwardt_2009}. The limb-darkening effect is modelled using a non-linear law given by the following equation: %~\ref{eq:4}, 
\begin{multline}\label{eq:4}
       I(\mu) = I_{0} \times [1-c_{1}(1-\mu^{\frac{1}{2}})-c_{2}(1-\mu) \\-c_{3}(1-\mu^{\frac{3}{2}})- c_{4}(1-\mu^{2})],
\end{multline}
where $\mu$=$\sqrt{1-x^{2}}$, $x$ is the normalised radial coordinate and I$_0$ is the star normalised flux. 
The coefficients (c$_{1}$,c$_{2}$,c$_{3}$,c$_{4}$), detailed in Table~\ref{table:2limb}, are computed using \citet{Sing_2010}, with the effective temperature, metallicity [Fe/H], and surface gravity given in Table~\ref{table:1parameters}. The UV spectral region contains strong stellar atomic lines that can behave anomalously at the limb, especially in temperature inhomogeneities. However, the present HST NUV light curves do not allow us to sample the ingress and egress portions of the transit precisely, so it is not straightforward to constrain the limb darkening coefficients to high accuracy. Besides, we tested other limb-darkening laws and found similar results to those presented below. \citet{Sing_2010} non-linear four parameters law is a good approximation for fitting HST NUV light curves.  
Transit parameters are initialised with values found in \citet{Brown_2017}. The limb-darkening coefficients used in this analysis are given in Table~\ref{table:2limb}.

\subsection{Spectral light curve fitting}
\label{subsec24:slcfitting}
We created a correction model for the white light curve using jitter correlation parameters and then applied this correction to all light curves obtained in various spectral wavelength ranges. The correction model is constructed by dividing the best-fit analytic white light curve model by the jitter parameters polynomial. We first divided the spectrum into several broadband spectral ranges ($\sim$200\text{\AA}), adopting the same selection as \citet{Sing_2019}. We also calculated higher resolution transmission spectra in 4\text{\AA} bins. The transit ($\sim$ 1\%) can still be resolved at this resolution except in spectral regions with strong stellar absorption and at the edge of spectral orders. We removed points where the transit was not detected.  
The planet-to-star radius ratio is a free parameter in the model, and its value is adjusted to obtain the best-fit spectrum. To validate the value found by the L-M method and estimate the 1-$\sigma$ error, we used the $\chi^2$ variation method \citep{Hebrard_2002} by varying the planet-to-star radius ratio and estimating the value of the $\chi^2$ as a function of the radius ratio. The minimum is found for the best-fit planet-to-star radius ratio, and the 2-$\sigma$ error bars are taken at $\rm\Delta\chi^2=4$, that is,
\begin{equation}
   \sigma = |\frac{R_{\rm P}/R_\star(\chi^{2}_{min}+4)-R_{\rm P}/R_\star(\chi^{2}_{min})}{2}|
.\end{equation}

\begin{figure*}[htpb]
\includegraphics[width=\textwidth]{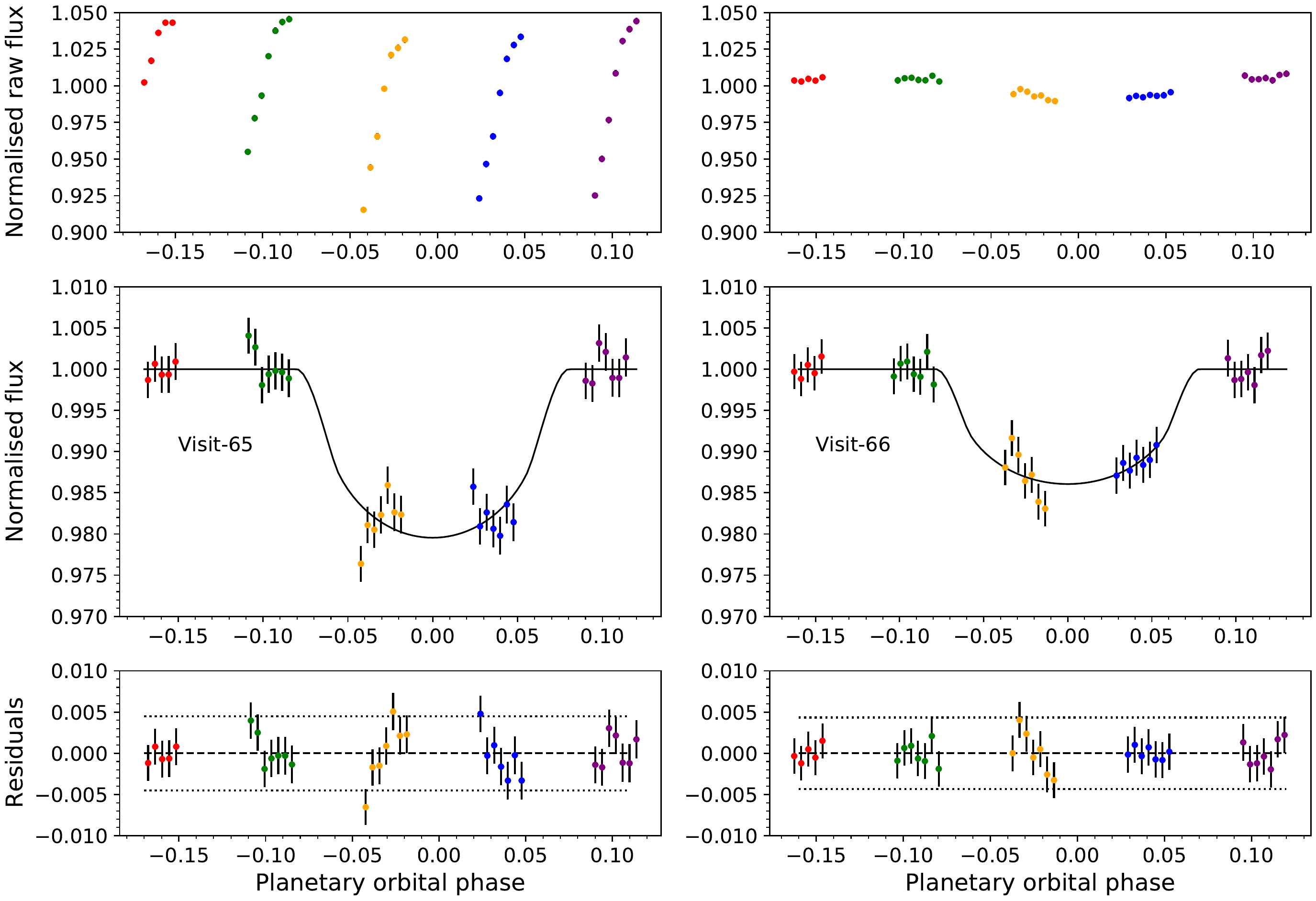}
\caption{WASP-79b white light curves for visit \#65 (left) and \#66 (right). Top: Normalised raw light curves. Middle: Flux corrected from systematic errors and fitted with a transit model. 
Bottom: Residuals between the flux corrected spectra, the best-fit models, and 2-$\rm \sigma$ error bars (dotted line). Visit \#65's raw data display a large ramp effect compared to visit \#66. This ramp is mainly correlated to the HST orbital phase.}
 \label{fig:1wlc}
\end{figure*}

\subsection{Validation using WASP-121b STIS data}\label{subsec25:validation}
To validate our fitting method described above, we applied it to WASP-121\,b's STIS data. WASP-121\,b is a hot Jupiter orbiting an F6V star with a similar temperature to that of WASP-79, which is 6460\,K \citep{Delrez_2016}. 
The planet has an inflated radius of 1.753\,R$_{\rm J}$ \citep{Bourrier_2020} and an equilibrium temperature reaching 2720\,K \citep{Mikal-Evans_2019}.  
These observations are also part of the PanCET program and were studied and published in \citet{Sing_2019}. 
Our analysis includes all HST NUV observations of WASP-121\,b, 
except for the first exposure of each orbit.
For WASP-121\,b, we find a R$_{\rm P}$/R$_{\rm \star}$ of $0.135\pm0.003$ for both visits, which is consistent within 1-$\rm \sigma$ with the values found by \citet{Sing_2019}, that is $0.1364\pm0.0110$ and $0.1374\pm0.0026$ for the first and second visit, respectively. We fitted $\sim $200\text{\AA} bins with the same systematic model found to fit the white light curve for the different visits optimally. We observed an increase in the wavelength-dependent R$_{\rm P}$($\rm \lambda$)/R$_{\rm \star}$ at a short wavelength with R$_{\rm P}$($2387\pm118$\text{\AA})/R$_{\rm \star}$ = $0.146\pm0.010$ for the first transit, which is within 1-$\sigma$ of the \citet{Sing_2019} result: $0.1530\pm0.0084$. We also obtained a 4\text{\AA} bin NUV transmission spectrum of WASP-121b using the second transit, which confirms the FeI, FeII, and MgII detections. For instance, we find R$_{\rm P}$/R$_{\rm \star}(2348\text{\AA})=0.292\pm0.053$,  
which is in the FeII absorption domain, and R$_{\rm P}$/R$_{\rm \star}(2796\text{\AA})=0.297\pm0.040$ in the MgII h-line of the doublet around 2800\,\AA . 
Therefore, we confirm that FeII and MgII are no longer gravitationally bound to the planet.

In conclusion, our WASP-121\,b analysis yields the same results as the ones found by \citet{Sing_2019}; this validates our procedure for the systematics correction and the planet radius estimates, as described above.

\section{Results}\label{sec3:results}

\subsection{White light curves fitting}\label{subsec:31wlc}
We applied our method to the STIS data obtained during the transit of WASP-79\,b. 
For each set of transit observations, we used all five~HST orbits but excluded the first exposure of each orbit. 
The two visits show highly different trends: the first visit \#65 presents more than 10\% variations in the normalised raw flux. This visit is highly affected by systematic effects, whereas the second visit, \#66, is of better quality with lower systematics (Fig.~\ref{fig:1wlc}). 

\subsubsection{First transit visit: \#65}\label{subsec:311wlc}

\begin{figure*}[htb]
  \centering
  \includegraphics[width =9cm]{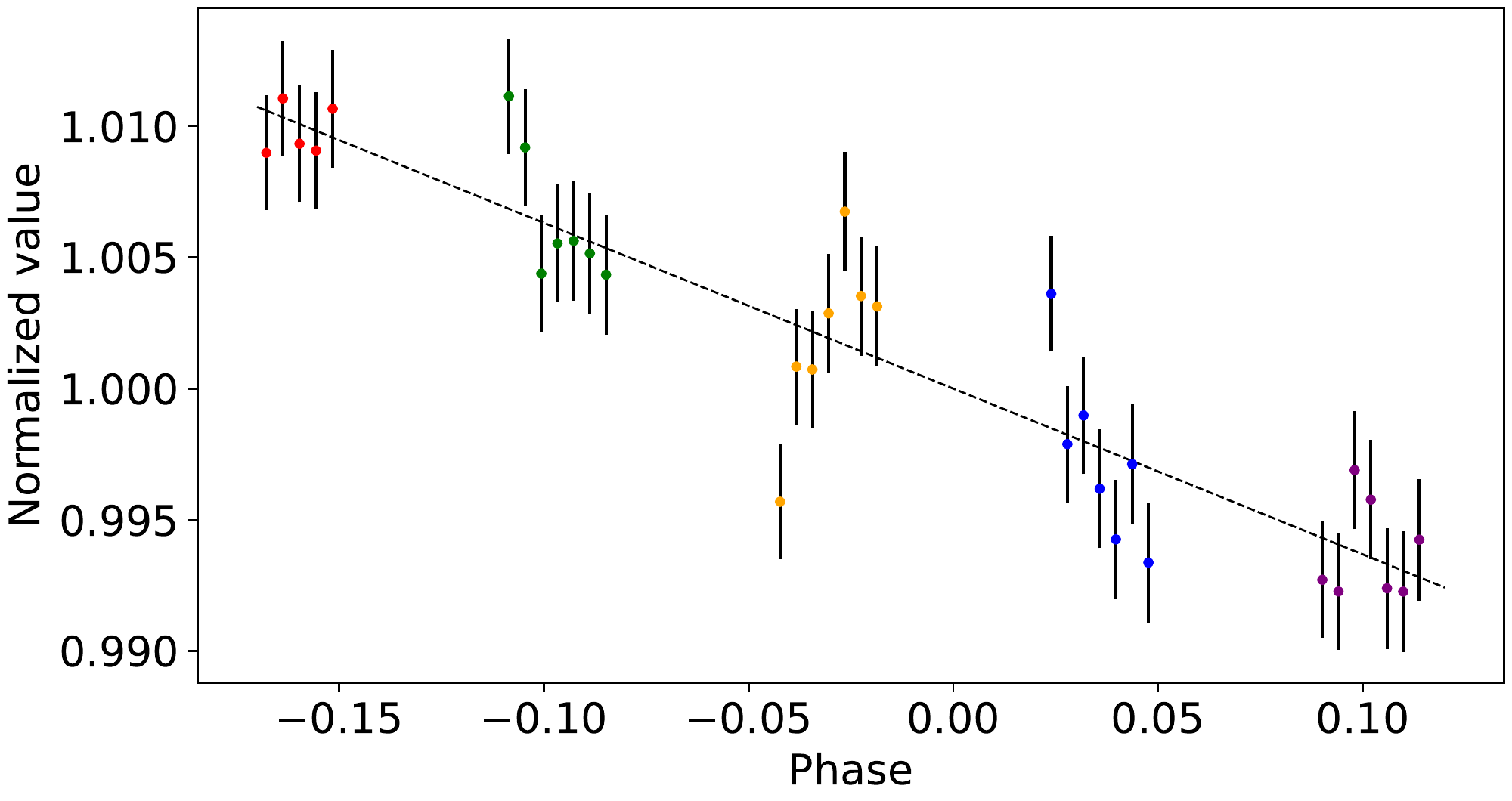}
  ~
  \includegraphics[width = 9cm]{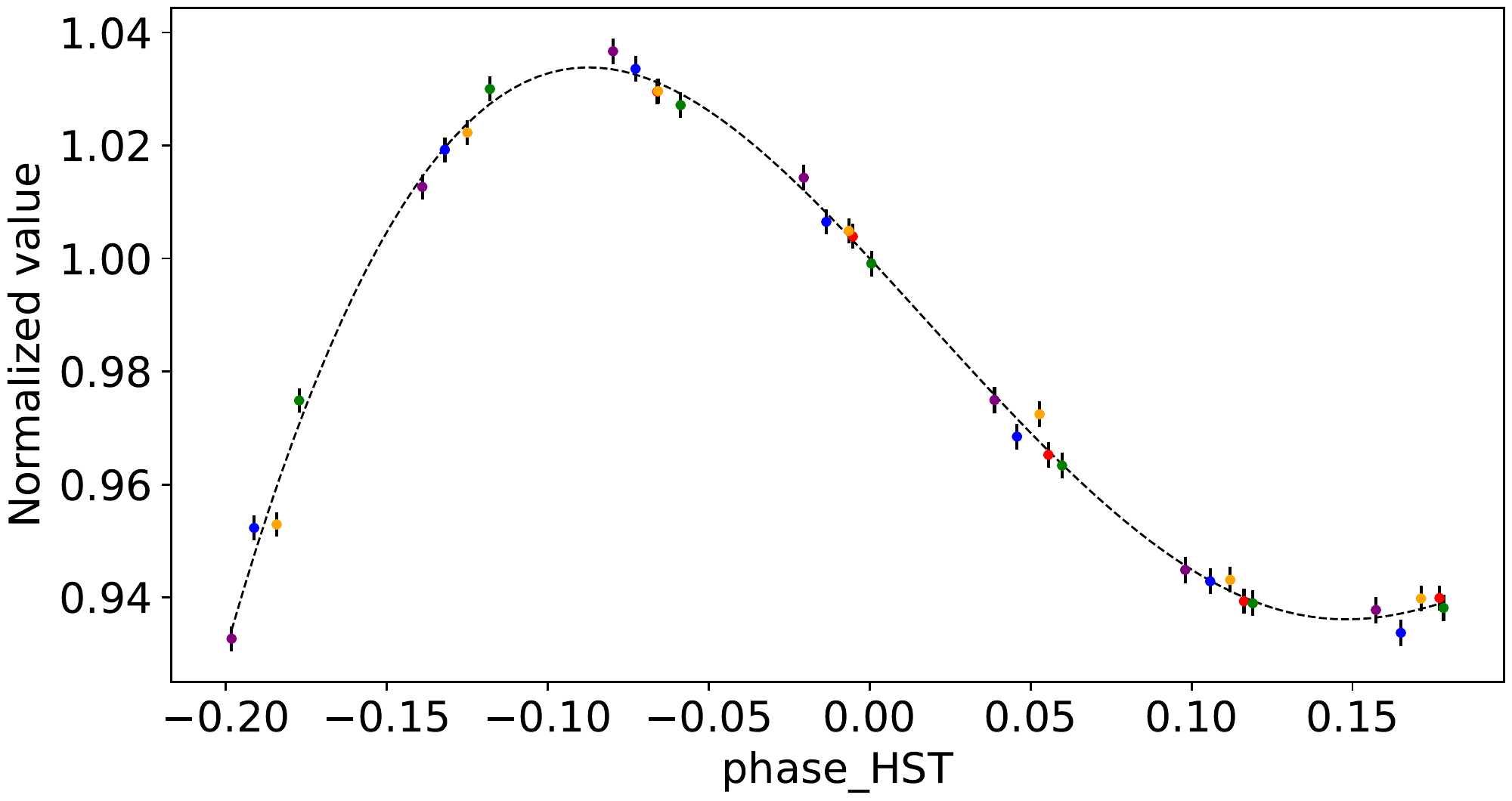} 
  
  \includegraphics[width = 9cm]{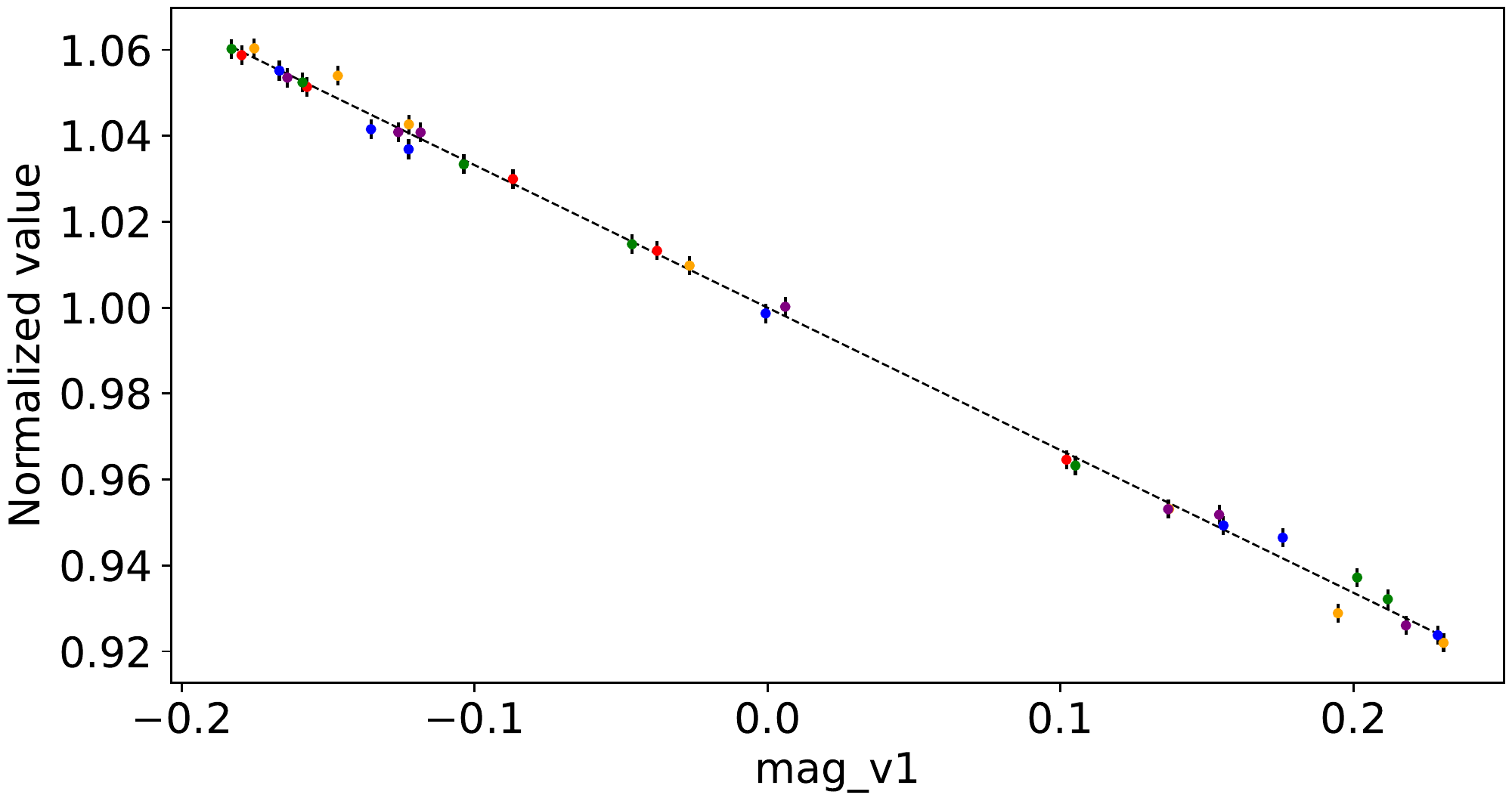}
  ~
  \includegraphics[width = 9cm]{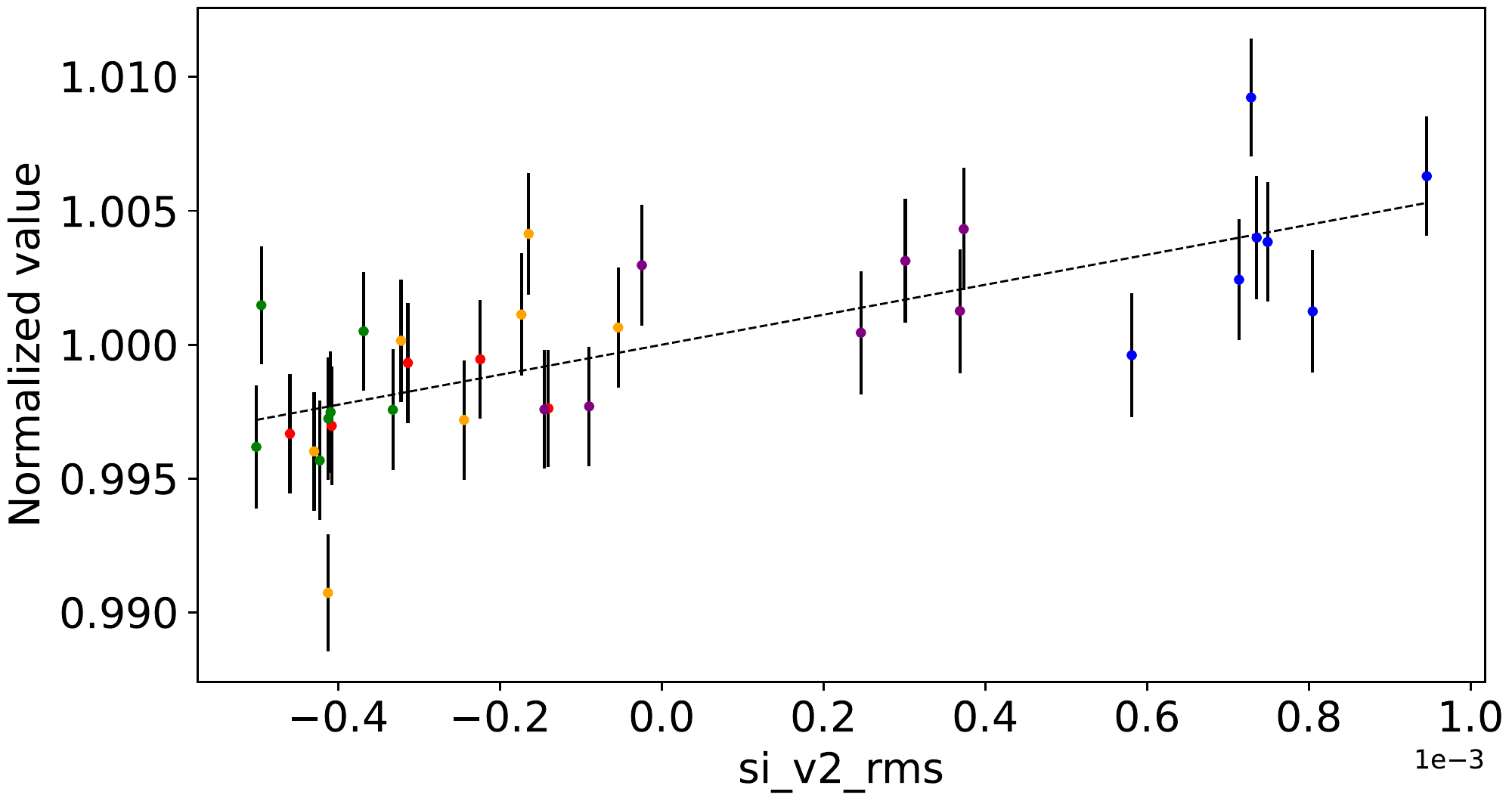}
  
  \includegraphics[width = 9cm]{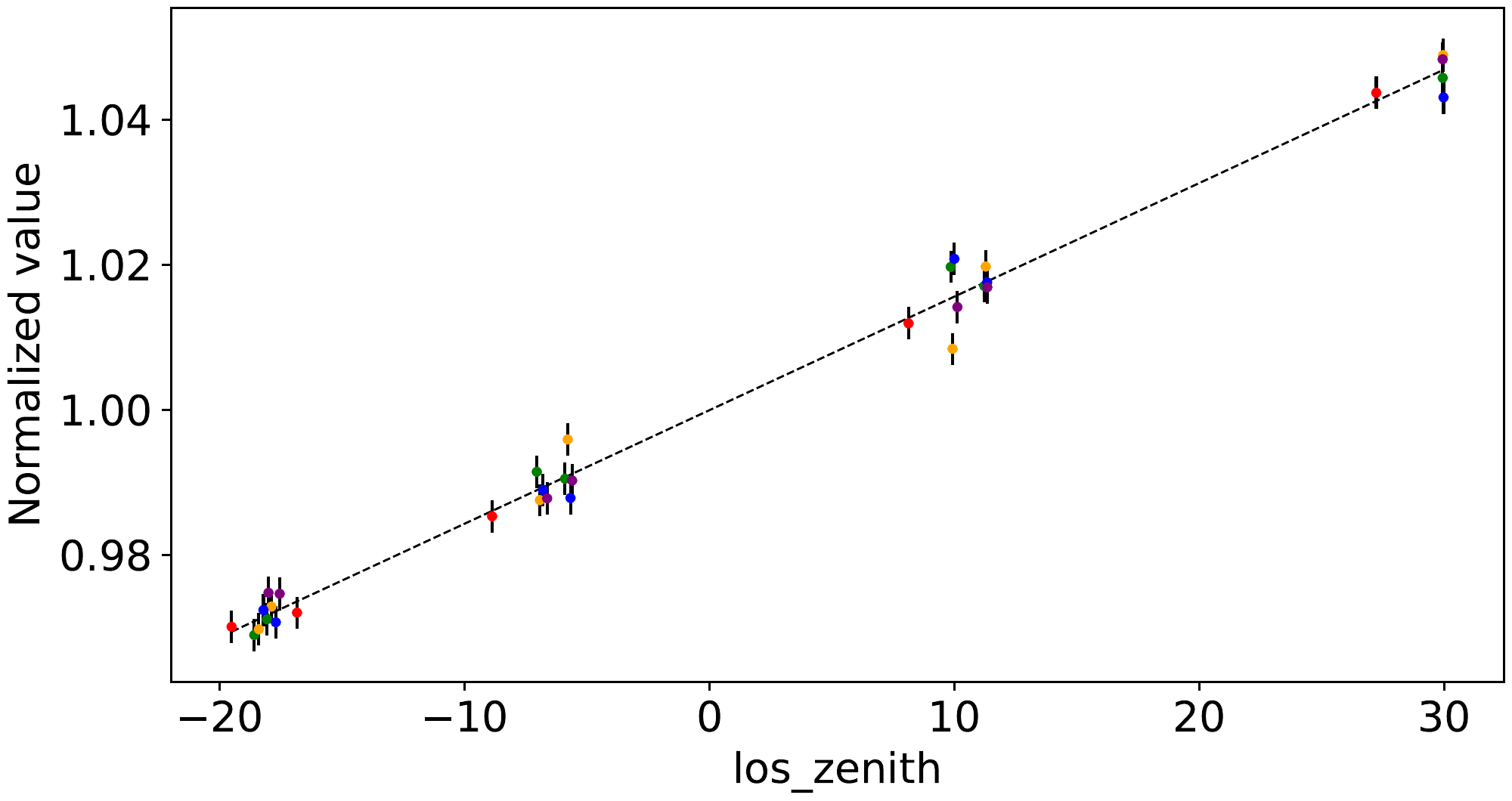}
  ~
  \includegraphics[width = 9cm]{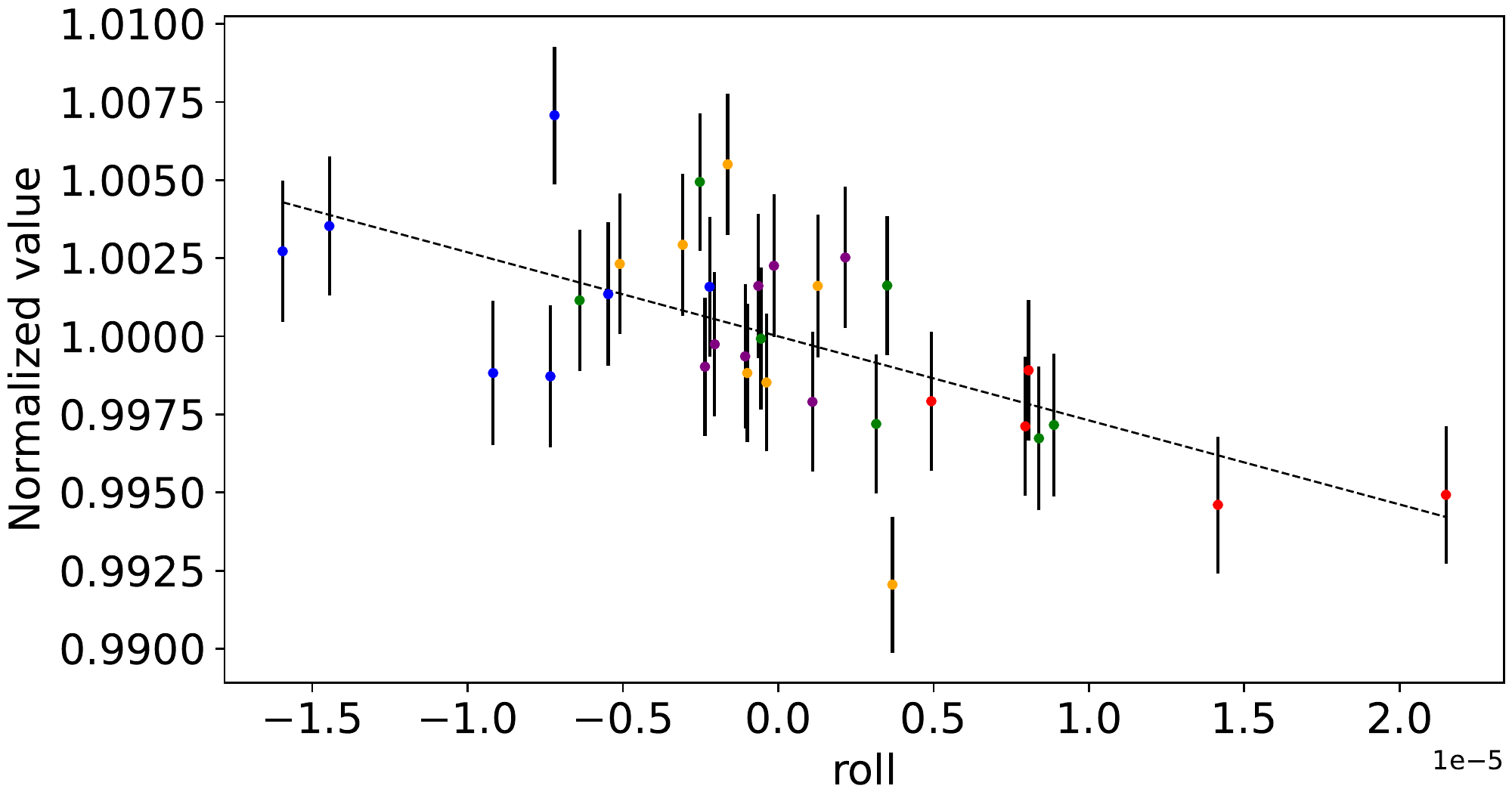} 
\caption{Polynomial fit (dotted line) included in correction model for white light curve of visit \#65 on WASP-79b. All polynomials are of degree one except for the polynomial with regards to the HST orbital phase, whose degree is eight.  }
 \label{fig:A2wlc65}
\end{figure*}

Following the procedure described in Sect.~\ref{sec:2data analysis}, we used a correction model to obtain the global fit. 
The data of visit \#65 requires a complex model with 15 free parameters in total: 
\begin{multline}
    S(X) = (1 + a_1\phi_t) \times (1 + a_2\phi_{HST} + a_3\phi_{HST}^2 + ...+ \\ a_9\phi_{HST}^8) \times (1 + a_{10} \text{ mag\_v1}) \times (1 + a_{11}\text{ si\_v2\_rms})  \\ \times (1 + a_{12}\text{ los\_zenith}) \times (1 + a_{13}\text{ roll}). 
\end{multline}
Correlations of the normalised raw flux with jitter parameters are plotted 
in Fig.~\ref{fig:A2wlc65}. 
The polynomial coefficients' values corresponding to the correction model's parameter are detailed in Table \ref{table:A1jit}. The analysis of the white light curve of the first visit, \#65, yields R$_{\rm P}$/R$_{\rm \star}$ = $0.1285\pm0.0021$. 
Even in the corrected light curve, some exposure points are outliers beyond 2-$\sigma$, and the light curve obtained within orbits~3 and~4 still presents systematic trends (see Fig.~\ref{fig:1wlc}). 
%Furthermore, the transit is fitted without egress or ingress. 
We performed a broadband analysis using the formalism of \citet{Sing_2019} and obtained the following values: R$_{\rm P}$/R$_{\rm \star}$($2400$\text{\AA}) = $0.1362\pm0.0063$, R$_{\rm P}$/R$_{\rm \star}$($2600$\text{\AA}) = $0.1254\pm0.0045$, R$_{\rm P}$/R$_{\rm \star}$($2800$\text{\AA}) = $0.1305\pm0.0035,$ and R$_{\rm P}$/R$_{\rm \star}$($3000$\text{\AA}) = $0.1255\pm0.0035$. This visit yields a deeper transit that is not compatible with previous studies. 

Therefore, the results of the first transit are most likely affected by strong instrumental systematics or suffer from stellar activity. 
Considering the large amplitude of the systematics in the raw measurements obtained during visit \#65 (see top left panel of Fig.~\ref{fig:1wlc}), we decided to focus our analysis on the transmission spectra obtained with visit \#66 and use the results from the observations of visit \#65 only for confirmation or consistency checks. 

Nonetheless, we note an increase in the transit depth at a low wavelength with a relative difference between the radius at 2400\text{\AA} and at 3000\text{\AA} of $\rm \Delta $R$_{\rm P}$/R$_{\rm \star}= 0.0107\pm0.0072$.
This consistent result gives confidence in visit \#66 data, as described in the following sections. 

\subsubsection{Second transit, visit \#66}
\begin{figure}[tb]
  \centering
  \includegraphics[width =\columnwidth]{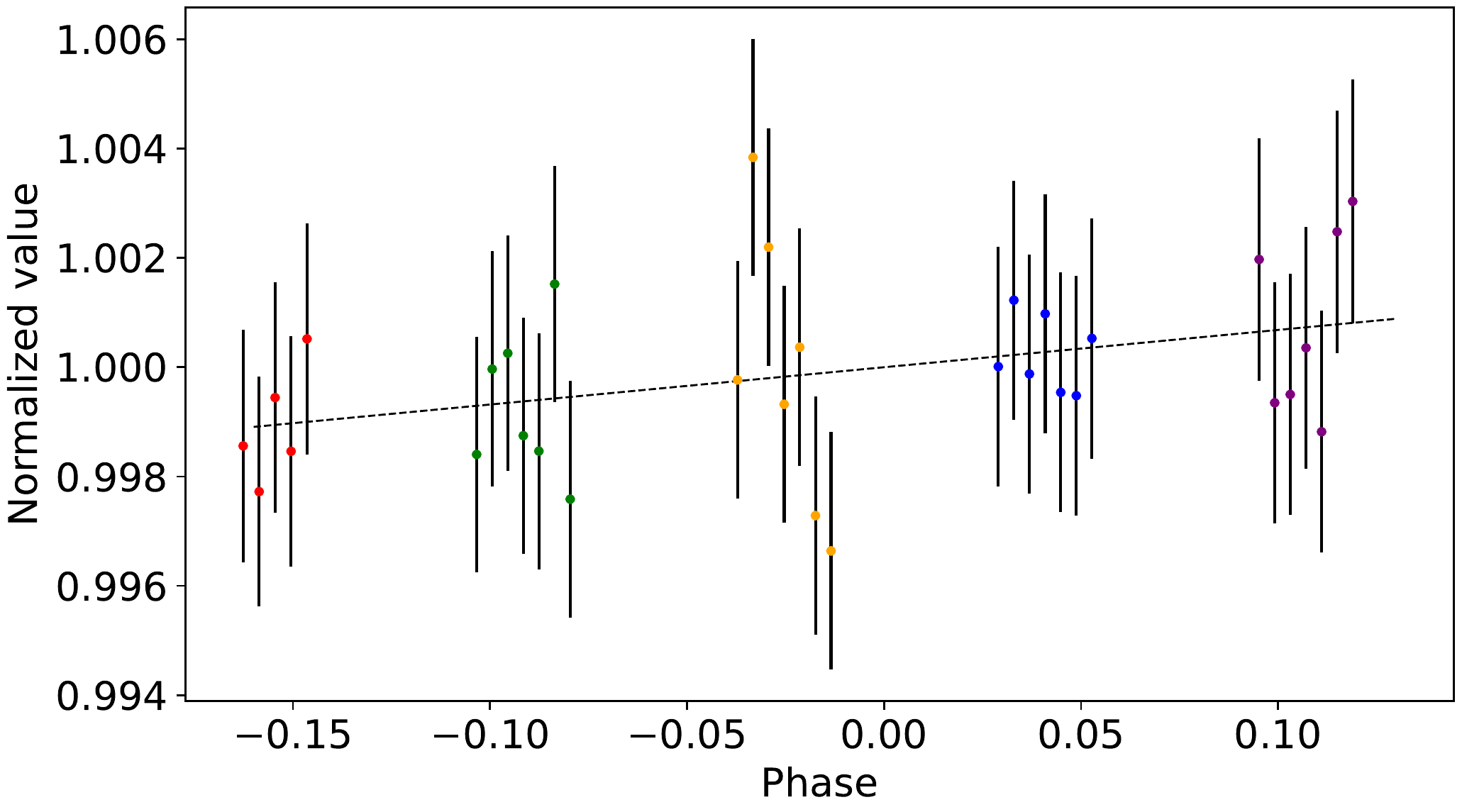}
  \includegraphics[width = \columnwidth]{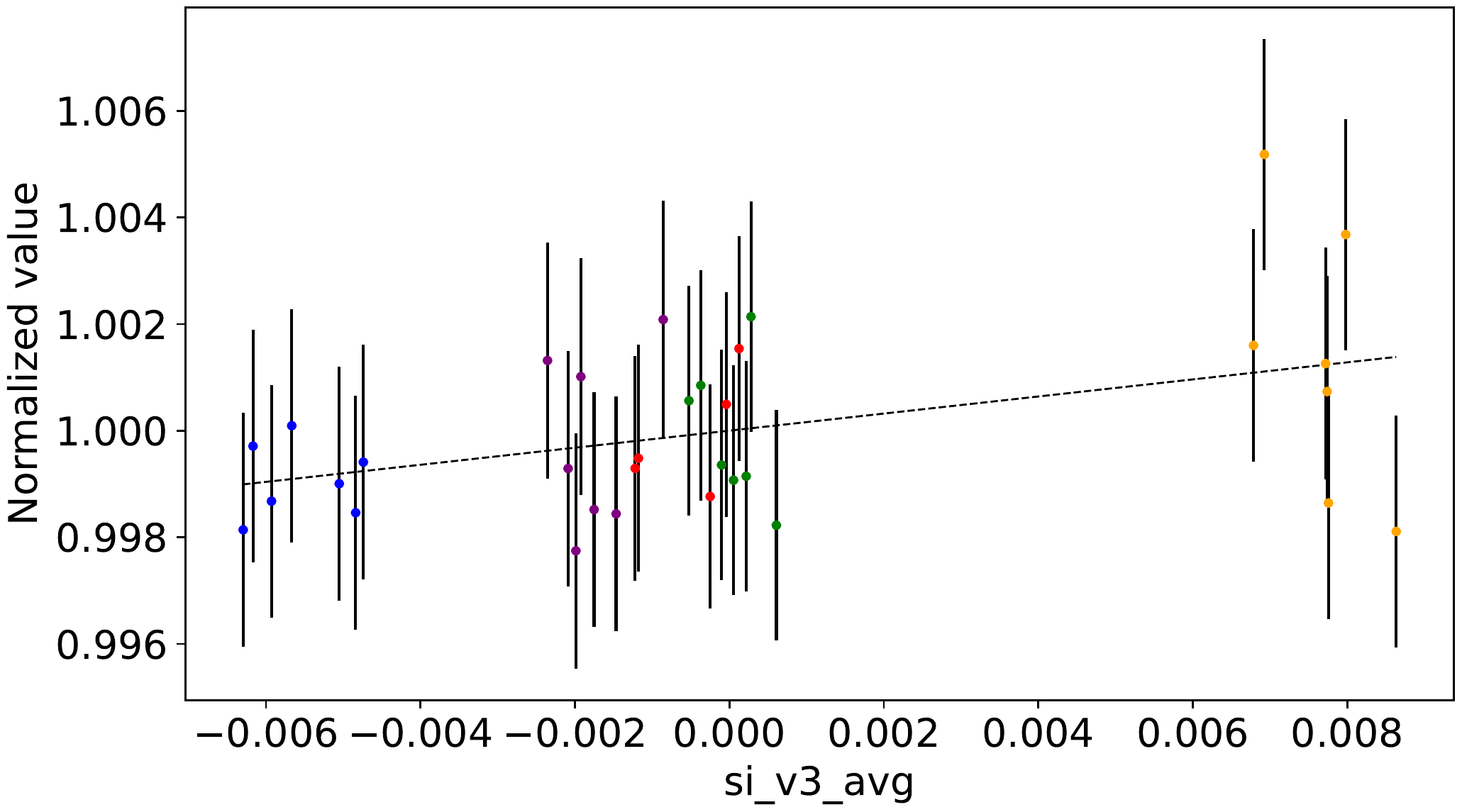}
\caption{Polynomial fit of degree 1 (dotted line) included in the correction model for the white light curve of visit \#66 on WASP-79b. }
 \label{fig:A3wlc66}
\end{figure}

The raw light curve of \#66 shows less systematics and can be fitted with only four~free parameters:
\begin{equation}
    S(X) = (1 + a_1\phi_t) \times (1 + a_2\text{ si\_v3\_avg})
\end{equation}
% \begin{multline}
%    S(X) = (1 + a_1\phi_t) \times (1 + a_2\text{ si\_v3\_avg})
%\end{multline}
The correlations of the normalised raw flux with jitter parameters are plotted 
in Fig.~\ref{fig:A3wlc66}. The value of the polynomial coefficients, including the jitter parameter {\tt si\_v3\_avg} (the mean jitter in V3 over 3 seconds), are in Table~\ref{table:A1jit}. Compared to visit\,\#65, visit\,\#66 presents a better transit phase coverage, and all residuals are below 2-$\rm \sigma$ (see Fig.~\ref{fig:1wlc}). The white light curve of the second visit, \#66, yields R$_{\rm P}$/R$_{\rm \star}$ = $0.1059\pm0.0025$.
Table~\ref{table:3litteratures_wlc} shows the different planet-to-star radius ratio measurements from previously published studies 
\citep{rathcke2021hst, Sotzen_2020}. Our new measurement is consistent with all previously published values for WASP-79b, 
particularly with the \citet{Brown_2017} value of $0.10440\pm0.00048$.

\begin{figure*}[htpb]
  \centering
  \includegraphics[width =\textwidth]{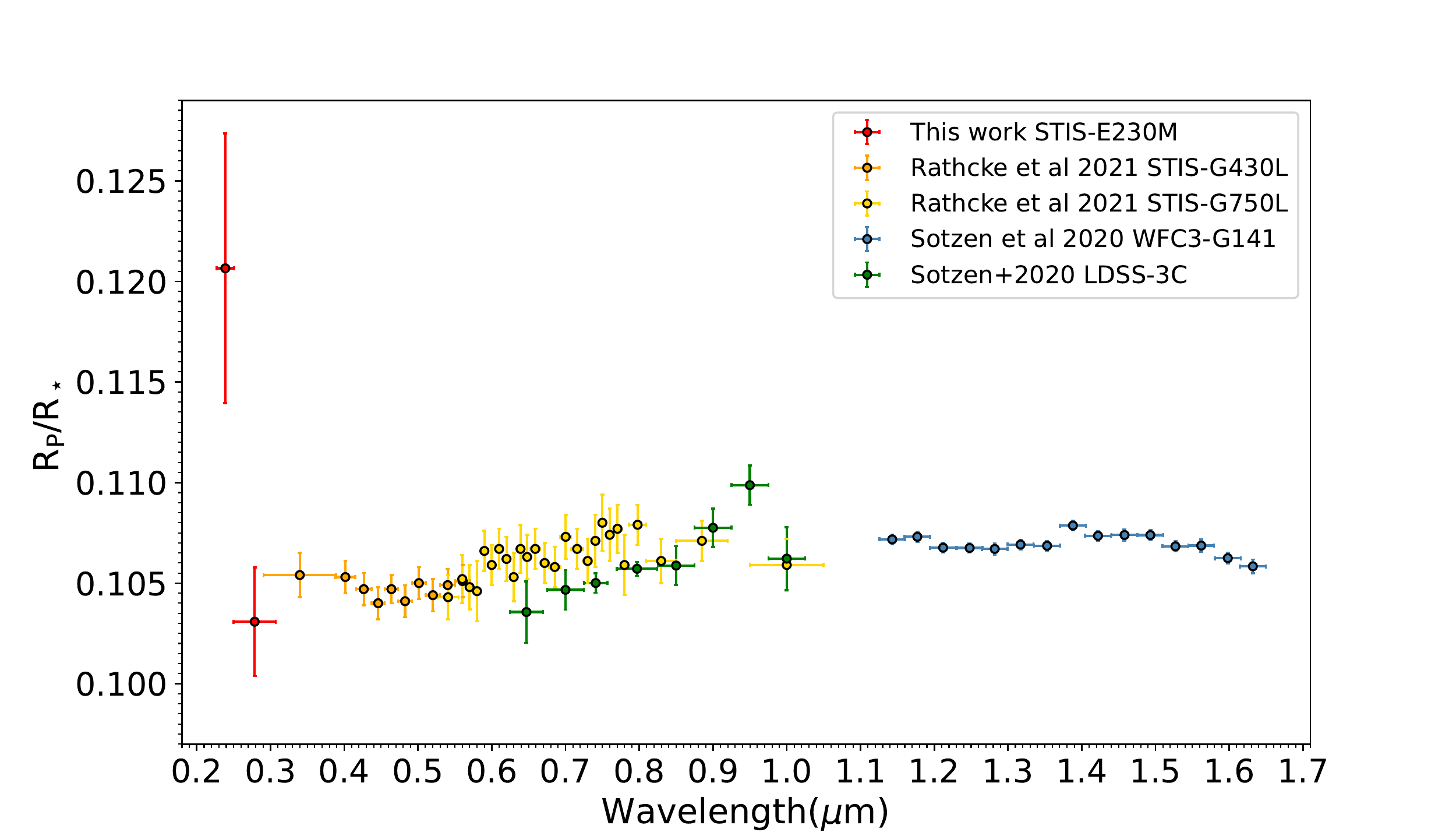}
\caption{WASP-79b HST transmission spectrum. NUV values are from the broadband analysis of visit \#66 obtained here 
using the STIS E230M observations (in red). For clarity, we represent two values around 2400\AA\ and 2700\AA\ (Bins 1 and 2 in Table~\ref{table:2broadband}). The optical values are from the analysis of \citet{rathcke2021hst} of STIS G430L and G750L data (orange and yellow). 
HST WFC3 observations in the NIR  are from \citet{Sotzen_2020} (blue).  }
  \label{fig:5HST_spectrum}
\end{figure*}

\subsection{Broadband analysis}\label{subsec:32broadband}

\begin{table*}[tb]
        \caption{Extracted planet-to-star radius ratio comparison with values found in the literature.}              
        \label{table:3litteratures_wlc}     
        %\centering                                     
        \begin{tabular}{l l l }          
                \hline\hline 
   Instrument & Bandpass &R$_{\rm P}$/R$_{\rm \star}$ \\\hline
    STIS E230M this work & 0.22-0.32 $\mu$m & 0.10590$\pm$0.0025 \\
    STIS G430L \citet{rathcke2021hst} & 0.29-0.57$\mu$m & 0.10519$\pm$0.00025\\
    STIS G750L \citet{rathcke2021hst} & 0.53-0.57$\mu$m & 0.10482$\pm$0.00040\\
    STIS G750L \citet{rathcke2021hst} & 0.59-1.02$\mu$m & 0.10662$\pm$0.00024\\
    TESS \citet{Sotzen_2020} & 0.59-1.02$\mu$m & 0.10675$\pm$0.00014\\
    LDSS-3C \citet{Sotzen_2020} & 0.60-1.0$\mu$m & 0.10782$\pm$0.00070 \\
    WFC3 G141 \citet{Sotzen_2020} & 1.10-1.70$\mu$m & 0.10621$\pm$0.00015 \\
    Spitzer \citet{Sotzen_2020} & 3.18-3.94 $\mu$m & 0.10594$\pm$0.00038\\
    Spitzer \citet{Sotzen_2020} & 3.94-5.06 $\mu$m & 0.10675$\pm$0.00048\\
                \hline
        \end{tabular}
\end{table*}

\begin{table}[tb]
        \caption{WASP-79~b radius measured in different broadbands.}             
        \label{table:2broadband}     
%       \centering                                     
        \begin{tabular}{l l l | c c }          
                \hline\hline 
                  & &  & \multicolumn{2}{c}{Visit \#66} \\
                 & $\lambda_C $(\text{\AA})& $\Delta\lambda $(\text{\AA}) &R$_{\rm P}$/R$_{\rm \star}$ & error \\    
                \hline 
        White & 2673 & 799 & $0.1059$ & $0.0025$ \\ 
        Bin 1 & 2786 & 572 & $0.1031$ & $0.0027$ \\
        Bin 2 & 2387 & 236 & $0.1207$ & $0.0067$ \\
        Bin 3 & 2600 & 200& $0.1072$ & $0.0051$  \\
        Bin 4 & 2800 & 200 & $0.1028$ & $0.0043$ \\
        Bin 5 & 2986 & 172 & $0.1016$ & $0.0042$ \\
        
        \end{tabular}
\end{table}

\begin{figure}[htpb]
  \centering
  \includegraphics[width =\columnwidth]{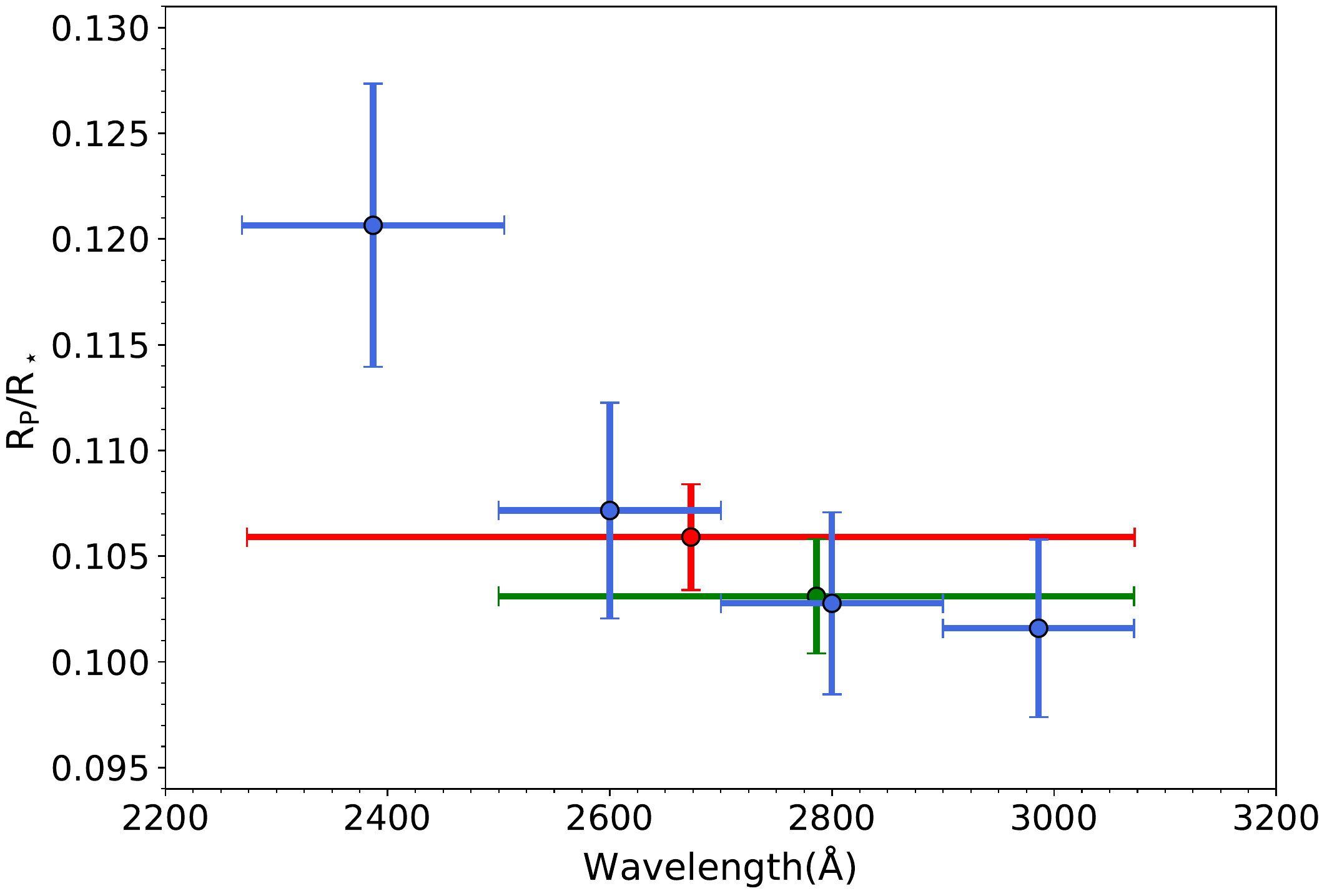}
\caption{Planet-to-star radius ratio measured in the broadband transmission spectra of visit \#66 (blue). 
The value from the white light curve is plotted in red, and the value from Bin 1 (2500 to 3100\AA) is in green.}
  \label{fig:2spectralbroad}
\end{figure}

We used the same broadband bin width of $\sim$200\text{\AA} as in  WASP-121b's STIS broadband analysis \citep{Sing_2019}, and we fitted the broadband data of WASP-79b using the same systematic detrending model used to correct the white light curve optimally. 
Broadband results are presented in Table~\ref{table:2broadband} and Fig.~\ref{fig:2spectralbroad}. We observe an increase in the planet-to-star radius ratio towards shorter wavelengths.
In the white light curve and the broadbands around 2600\,\AA , 2800\,\AA,\ and 2900\,\AA , the planet-to-star radius ratio is found to be compatible, within 1-$\sigma,$ with the findings of \citet{rathcke2021hst} at 0.3$\mu$m. 
However, at shorter wavelengths around 2400\,\AA , the ratio R$_{\rm P}$/R$_{\rm \star}$(2400\AA )=0.1207$\pm$0.0067 is found to be significantly higher. For this reason, we computed the planet-to-star radius ratio for the 2500-3000\AA\ band, where no variation in wavelength is detected. 
We found R$_{\rm P}$/R$_{\rm \star}$=0.1031$\pm$0.0027. 

The relative difference between the planet-to-star radius ratio at 2400\,\text{\AA} and 3000\,\text{\AA} is $\rm \Delta $R$_{\rm P}$/R$_{\rm \star}=0.0191\pm0.0079$. This result is consistent within 1-$\sigma$ with the value of $0.0107\pm0.0072$ found using visit \#65 data in Sect.~\ref{subsec:311wlc}. 
This increase in the absorption depths at short wavelengths can be explained by the absorption of heavy ionic or atomic species \citep{Lothringer_2020} or by the presence of hazes in the upper part of the atmosphere. 
In short, we observe a significant increase in 
the apparent radius of the planet at shorter wavelengths 
that could be due to the presence of clouds or hazes. 

Figure~\ref{fig:5HST_spectrum} shows the overall HST transmission spectrum of WASP-79\,b, including our NUV measurements obtained with the data of the second visit \#66, 
\citet{Sotzen_2020} measurements in the NIR (HST WFC3) and \citet{rathcke2021hst} measurements in the NUV and visible wavelength range (HST STIS G430L and G750L). We represent the value around 2400\AA\ in a 200\AA\ band and the large band value after 2500\AA\ to show the steep increase in the planet-to-star radius ratio at short wavelengths. 

Combining different transmission spectroscopy datasets is difficult, even more so when there is no spectral overlap. The orbital parameters and limb-darkening coefficients can differ from one study to another. The treatment of systematic effects and stellar activity can also vary, leading to variations in planet-to-star radius ratio and transit depth measurements \citep{Tsiaras_2018, Yip_2020, Changeat_2020, Pluriel_2020, Edwards_2021}. Even while using the same system parameters, prescriptions for limb darkening and a lack of stellar activity, \citet{Nikolov_2013} already highlighted differences in absolute radius level for HAT-P-1b when combining STIS with WFC3. 
We decided to keep the values only from the HST instruments even though \citet{Sotzen_2020} and \citet{rathcke2021hst} showed 
the compatibility of LDSS-3C transmission values on WASP-79\,b in the optical and NIR. 
Our measurement in the NUV obtained using the white light curve of the visit \#66 is compatible, within 1-$\sigma,$ with all other 
transmission spectra. Nonetheless, it shows a trend of increasing absorption towards shorter wavelengths.

\subsection{Narrow-band analysis}
\label{subsec:33spectral}

\begin{figure} [htb]
  \centering
  \includegraphics[width =\columnwidth]{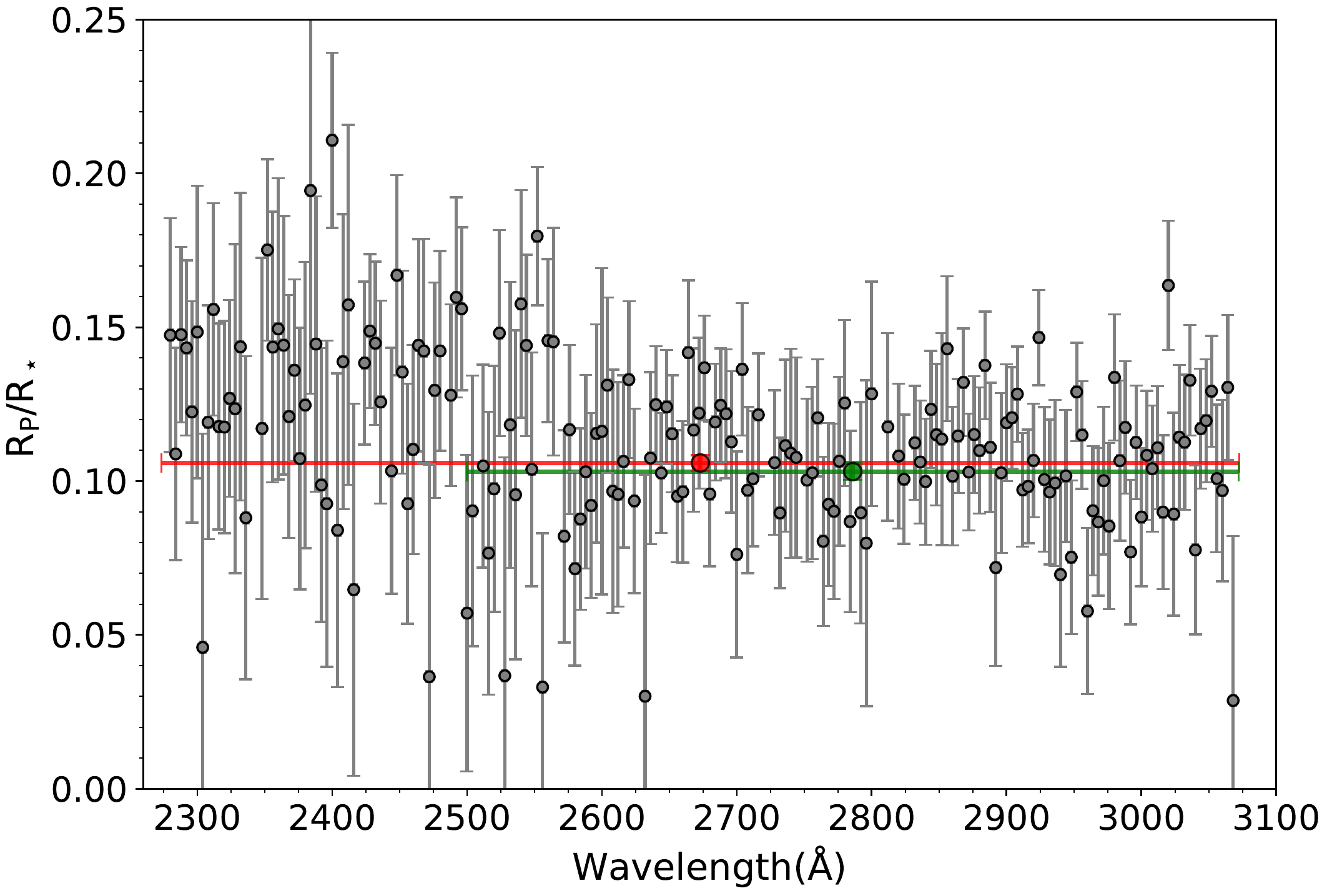}
\caption{WASP-79b NUV transmission spectra in 4\text{\AA} bins for visit \#66. We indicate the overall NUV transmission spectrum (white light curve) R$_{\rm P}$/R$_{\rm\star}$(NUV) = 0.1059 in red, along with the planet-to-star radius ratio after 2500(\AA) in green R$_{\rm P}$/R$_{\rm\star}$(>2500\AA) = 0.1031. }
  \label{fig:3spectralbins}
\end{figure}

To investigate the possibility of the presence of heavy species in the upper atmosphere producing a dense forest of narrow absorption lines, we calculated the transmission spectrum in narrow bands of 4\text{\AA} width (Fig.~\ref{fig:3spectralbins}). 
We search for an increase in the apparent radius of the planet at specific wavelengths that could be due to the presence of heavy metal species at very high altitudes or escaping the atmosphere. 
The wavelengths of the highest absorption in the transmission spectrum can be seen in Fig.~\ref{fig:3spectralbins}, some exceeding 0.18 in planet-to-star radius ratio. None of them corresponds to known metallic species absorbing in this part of the spectrum. We note that the value at 2384\text{\AA} could correspond to an Fe II line, usually found at 2382\text{\AA}. However, there is no detection of excess absorption of Fe II in other lines with similar oscillator strengths. 
We note that the data reduction process described in Sect.~\ref{sec:2data analysis} does not converge in the vicinity of two strong FeI lines (2484 and 2719\text{\AA}). 
This is likely due to the very low flux level in the middle of the line because of the stellar atmosphere absorption. 

The spectrum does not provide clear evidence of Fe I or Fe II absorption that could have explained the observed increase in the apparent radius at short wavelengths.
Nonetheless, the absorption spectrum in narrow bands confirms the global increase in the apparent radius of the planet at short wavelengths. 
The planet-to-star radius ratio weighted average is $0.1233\pm0.0052$ below 2500\text{\AA} and $0.1022\pm0.0021$ beyond. This simple computation compares nicely to the broadband analysis and the values found around 2400\text{\AA}, R$_{\rm P}$/R$_{\rm\star}$(2400\AA) = 0.1207$\pm$0.0067, and after 2500\text{\AA}, R$_{\rm P}$/R$_{\rm\star}$(>2500\AA) = 0.1031$\pm$0.0027. We also computed each bin's transmission spectrum shifted by 2\text{\AA}. This analysis confirmed the shape of the spectrum.

\section{Discussion}\label{sec:4discussion}

\subsection{Scale height of the atmospheric absorption}

At the shortest wavelengths ($\sim$2400\AA ), the planet's radius measured in the broadband and the narrow band is significantly higher than at longer wavelengths ($\ge 2600$\AA ).
Although this increase in planetary radius is consistently observed in the two visits, \#65 and \#66, here we consider only the quantitative estimates obtained with the \#66 visit, which provided better quality data with fewer systematic errors.  

In the broadband at 2400\,\AA,\, we obtain R$_{\rm P}$/R$_{\rm \star}$=0.1207$\pm$0.0067, which is about 16\% bigger than the planet size as seen in the optical. 
From 3000\,\AA\ to 2400\,\AA, the increase in planet size is measured to be 
$\rm \Delta $R$_{\rm P}$/R$_{\rm \star}=0.0191\pm0.0042$. We performed a joint fit of the two light curves to accurately compute the difference between the planetary radius and the uncertainty. We concatenated the two light curves into one matrix and adjusted each light curve with the same correction model as for the broadband analysis. However, instead of fitting for the two planet-to-star radius ratios in the two separate bins, we fitted for the planet-to-star radius ratio of the first bin and the $\Delta $R$_{\rm P}$/R$_{\rm \star}$. We then computed the uncertainty using the variation of the $\chi^2$ described above. This method is justified to find the increase in planetary size and absorption as we are looking for a relative measurement, not two independent, absolute planetary radii. While computing the two radii individually and taking the difference, we propagate the uncertainty on the coefficients of the correction model; the joint fit cancels the sharing part of the uncertainty of the systematic effects. The error on the relative difference of radii is less than the sum of the errors on the absolute radii because the systematic errors are the same for both radii, and the shared part of the systematic errors for both radii in the free parameters have the same values in the new simultaneous fit. Our finding corresponds to an increase in radius shortward of 2400\,\AA to a 4.5-$\sigma$ effect, which rules out a statistical fluctuation. 

The increase in the apparent planet size at specific wavelengths is commonly interpreted as extra absorption in the atmosphere. 
Here, the increase of 16\%\ in  the size of the planet is so large that the variation of the planet's gravity reaches 30\%  
between the bottom atmosphere and the altitude at which the atmosphere is optically thick at 2400\AA; the usual derivation of the atmospheric scale height must be adapted.

With a constant gravity, the hydrostatic equilibrium equation 
is $dP/P = -(\mu g  / kT) dr$, where $P$ is the pressure, $\mu$ the mean molar mass of the atmosphere, $g$ the gravity, $T$ the temperature, and $r$ the distance to the planet centre. This allows us to define the atmospheric scale height: $H=kT/\mu g$. The pressure $P$ as a function of the altitude $z$ is then $P(z)=P_0 \exp (-z/H)$, where $P_0$ is the pressure at zero altitudes. 
However, with an atmospheric thickness that is not negligible compared to the planet's radius, we need to consider the planet's gravity decrease with altitude. The new hydrostatic equilibrium equation is $dP/P = -(\mu G M_p  / kT r^2) dr$, where $M_p$ is the planet's mass and $G$ is the gravitational constant. The pressure vertical profile becomes $P(z)=P_0 \exp (-z R_p/H (R_p+z))$ or $P(z)=P_0 \exp (-z /H^\prime)$, where the modified altitude dependent scale height is $H^\prime(z) = H\cdot(R_p+z)/R_p$. 

For WASP-79\,b, we have a temperature $T=1716\pm 25$\,K and a planet gravity $g=GM_p/R_p^2=10.6\pm1.6$\,ms$^{-2}$ \citep{Brown_2017}. Assuming an hydrogen-helium atmosphere with $\mu =2.3$, this yields 
a scale height of $H=580\pm 100$\,km, and a ratio $H/R_{\rm \star}=5.5\times 10^{-4} \pm 1.0 \times 10^{-4} $. 
Finally with an atmospheric absorption thickness of $\rm \Delta $R$_{\rm P}$/R$_{\rm \star}=0.0191\pm0.0042$, we find a ratio of 
$\Delta R_{\rm P}/H \approx 34$.
If we consider the gravity variation with altitude, this ratio is decreased to $\Delta R_{\rm P}/H^\prime \approx 29$.
This value remains huge; moreover, the absorption at a high altitude, as observed at 2400\,\AA,\, takes place at a pressure that is $e ^{-29}=2\times 10^{-13}$ lower than at the altitude where the atmosphere is optically thick at $\sim 3000$\,\AA . 

We identified two other processes that could increase the scale height and thereby require less strong absorption than computed above. First, the scale height value uses a calculated equilibrium temperature, not a measured temperature. Spitzer secondary eclipses of this planet give a day side temperature of about 1950$\pm$85K \citet{Garhart_2020}. The planet is irradiated strongly enough that it may not redistribute heat efficiently, and the limb temperatures may be similar to the Spitzer day-side temperatures. However, we re-computed the scale height using the day-side temperature, and we find $H=660\pm 130$\, km, which translates to a ratio of $\Delta R_{\rm P}/H \approx 30$. We find a ratio of $\approx 25$ while considering the gravity variation with altitude, which remains very large. The impact of the temperature is marginal. Besides, the mean molecular weight could be lower than 2.3 amu due to hydrogen dissociation. Even if we consider that the hydrogen molecules may be partially dissociated, leading to a smaller mean molar mass ($\mu$\,$\sim$1), we find $\Delta R_{\rm P}/H^\prime \approx 13$ with a temperature of 1716K and $\approx 11$ with a temperature of 1950K, and the conclusion remains the same. Although identifying the main absorber at $\sim$2400\,\AA\ remains a puzzling task, it must have an extremely large cross-section to be optically thick at such a low density; clouds and hazes appear to be the most plausible carrier of the detected absorption at the shortest wavelengths.

\subsection{Comparison with the Roche-lobe equivalent radius }\label{subsec:41RL}

\begin{figure}[htb]
  \centering
  \includegraphics[width =\columnwidth]{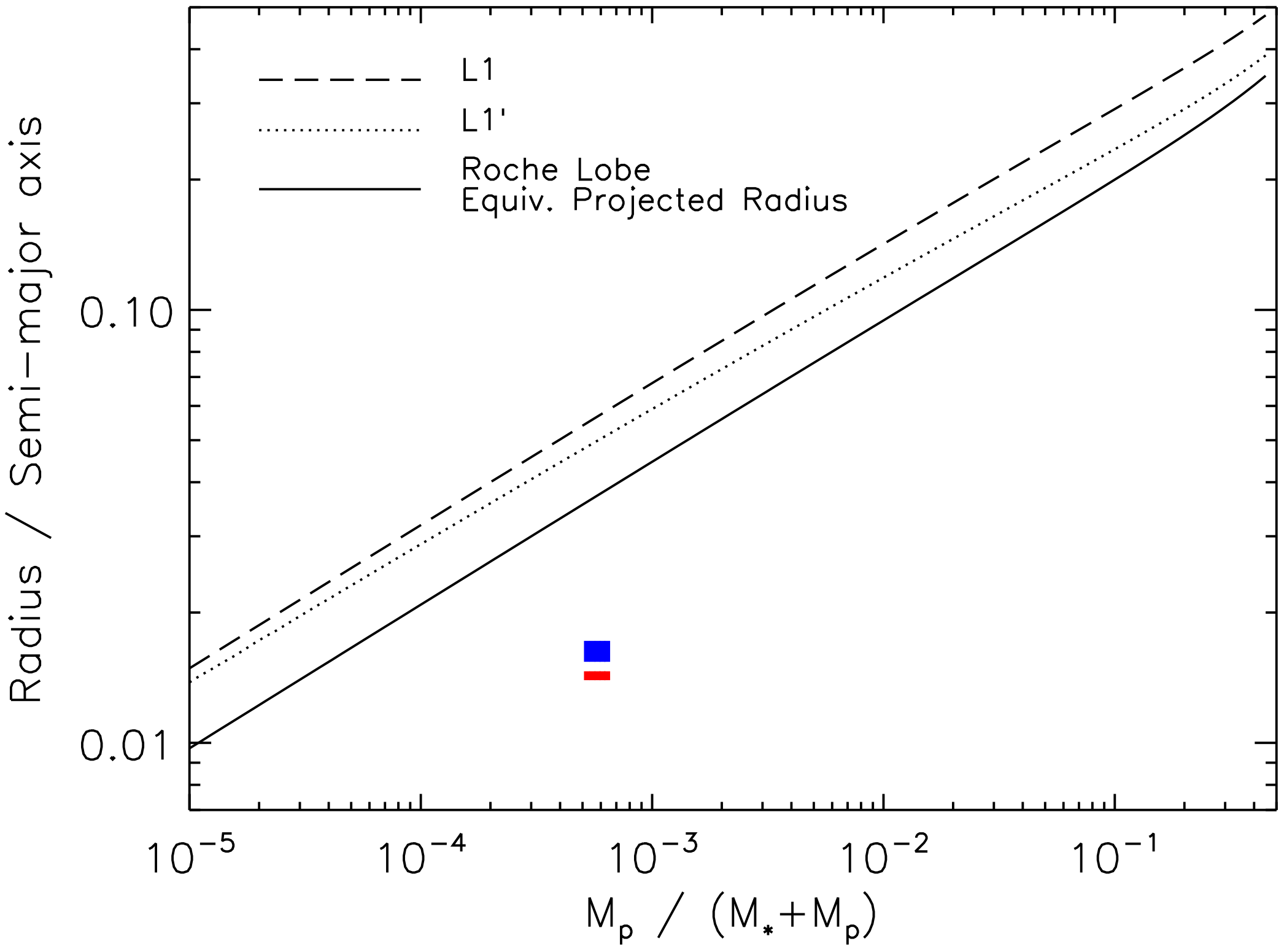}
\caption{Ratio of Roche-lobe size to the orbital semi-major axis as a function of the planet-to-star mass ratio. The size of the Roche lobe can be calculated using the distance of the planet centre to the L1 point (dashed line), to the L1$^\prime$ point (dotted line), or by calculating the projected size of the Roche lobe occulting the stellar disc during a transit observation. The latter is about 2/3 of the size calculated by considering the L1 or L1$^\prime$ points. The measurements for WASP-79~b in the NUV white light curve and at 2400\AA\ are plotted in red and blue, respectively.}
  \label{fig:rochelobe}
\end{figure}

The high thickness of the atmosphere detected at 2400\,\AA\ raises the question of possible geometrical escape as defined by \cite{Lecavelier_2004}, which occurs when the Roche lobe is filled up with the upper atmospheric gas.
Thus, the high altitude of absorbers detected at the shortest wavelengths needs to be compared to the size of the Roche lobe. 
For that purpose, we calculated the distance of the L1 and L1$^\prime$ Lagrange points of the equipotential surface 
to the planet centre as a function of the planet-to-star mass ratio (Fig.~\ref{fig:rochelobe}). 
We also calculated the equivalent radius of the occulted area during a Roche lobe transit. 
Because of the elongated shape of the Roche lobe, this equivalent radius is about 2/3 
of the distance between L1 and the planet centre \citep{Vidal_Madjar_2008, Sing_2019}. 

For WASP-79\,b, we found that with a planet-to-star mass ratio of $5.8\times 10^{-4}$ 
and a semi-major axis-to-star radius ratio of $a/R_\star=7.407$, 
the transit of the Roche lobe corresponds to an occultation by a disc with a radius of 0.276 times 
the radius of the star. 
The measured radius at 2400\,\AA\ is only 44\% this size. Even the highest values in the narrow band spectrum correspond to about 76\% of the size of the Roche lobe. Therefore, none of the absorption depths 
measured in the spectrum of WASP-79\,b reach the absorption that would be caused by an optically thick Roche lobe, and we are left to consider that the detected absorptions are 
due to components at high altitudes of the upper atmosphere.

None of the absorption features exceed the theoretical Roche-lobe radius. There is no evidence of atmospheric hydro-dynamical escape in our NUV transmission spectrum.

\subsection{Faculae}

Data analysis from \cite{rathcke2021hst} of WASP-79 b showed that the spectrum from 0.3 to 1.0$\mu$m is significantly affected by the presence of facul\ae\ on the stellar surface. 
They found that about 15\% of the stellar photosphere is covered 
by facul\ae\ that is $\sim$500\,K hotter than the mean temperature of the star. Because of their different blackbody temperatures, facul\ae\ and spots modify the planet-to-star radius ratio measured through transit observations, even if the planet does not pass in front of these features \citep{Pont_2013, McCullough_2014}.

In the presence of unocculted stellar spots or facul\ae, 
the measured radius ratio (R$_{\rm P}$/R$_{\rm \star}$)$_{\rm mes}$ is given by 
$$
\left(\frac{R_{\rm P}}{R_{\rm \star}}\right)_{\rm mes} =
\left(\frac{R_{\rm P}}{R_{\rm \star}}\right)_{\rm real} 
/
\sqrt{1-f\cdot\left(1-\frac{F_{\rm spot} (\lambda)}
{F_{\rm \star} (\lambda)}\right),}
$$
where (R$_{\rm P}$/R$_{\rm \star}$)$_{\rm real}$ is the real physical radius ratio, $f$ is the fraction of the stellar area covered by the spots or facul\ae , 
and $F_{\rm spot} (\lambda)$ and $F_{\rm \star} (\lambda)$ are the specific intensities of the spots (or facul\ae ) and the star, respectively. Assuming a blackbody at 6600\,K, we have
$F_{\rm \star}$(2300\AA )=
$1.4\cdot 10^{7}$\,W\,m$^{-2}$\,str$^{-1}$\,$\mu$m$^{-1}$,
and
$F_{\rm \star}$ (3000\AA )=
$3.4\cdot 10^{7}$\,W\,m$^{-2}$\,str$^{-1}$\,$\mu$m$^{-1}$.
With a 500\,K higher temperature of 7100\,K for the facul\ae , we have
$F_{\rm spot}$(2300\AA )=
$2.76\cdot 10^{7}$\,W\,m$^{-2}$\,str$^{-1}$\,$\mu$m$^{-1}$,
and
$F_{\rm spot}$(3000\AA )=
$5.72\cdot 10^{7}$\,W\,m$^{-2}$\,str$^{-1}$\,$\mu$m$^{-1}$.
Finally, with $f=0.15$, we obtain that the measured radius ratio is larger than the real ratio by a factor of 1.07 at 2300\AA\ and 1.05 at 3000\AA . 
Therefore, the planet-to-star ratio increases towards shorter wavelengths between 3000\AA\ and 2300\AA\ due to the facul\ae\ being only about 2\%. 
Even if we consider the extreme case of the error bars given by \cite{rathcke2021hst}, that accounts for 25\% of the stellar surface covered by facul\ae\ with $\Delta T$=900\,K. We found an increase of only 6\%.
To reproduce the observed $\sim$20\% increase in planet ratio, the facul\ae\ should have a brightness temperature of at least 9\,000\,K over 15\% of the stellar disc. Even if a blackbody is a poor approximation of the facul\ae\ in the UV, unrealistic brightness would be required to explain the observations.
Although the unocculted facul\ae\ have some effect on the measured radius ratio, this effect is negligible. It does not explain the observed amplitude of the increase in the radius ratio towards shorter wavelengths.  

\subsection{1D and 2D atmospheric simulations}\label{subsec:421Dsimulations}
The NUV transmission does not show evidence of photo-evaporation. However, it presents high atmospheric features proving the presence of clouds, hazes, or atomic species at very high altitudes in the atmosphere of WASP-79\,b, while optical observations \citep{Sotzen_2020, Skaf_2020} using HST WFC3 suggest the presence of H$_2$O and FeH in deeper layers. We decided to compare the observations to the predicted atmospheric composition and temperature profile by modelling the planet's interior using the Exoplanet Radiative-convective Equilibrium Model (Exo-REM) \citep{Baudino_2015, Charnay_2018, Blain_2021}.
\begin{figure}[htpb]
  \centering
  \includegraphics[width =\columnwidth]{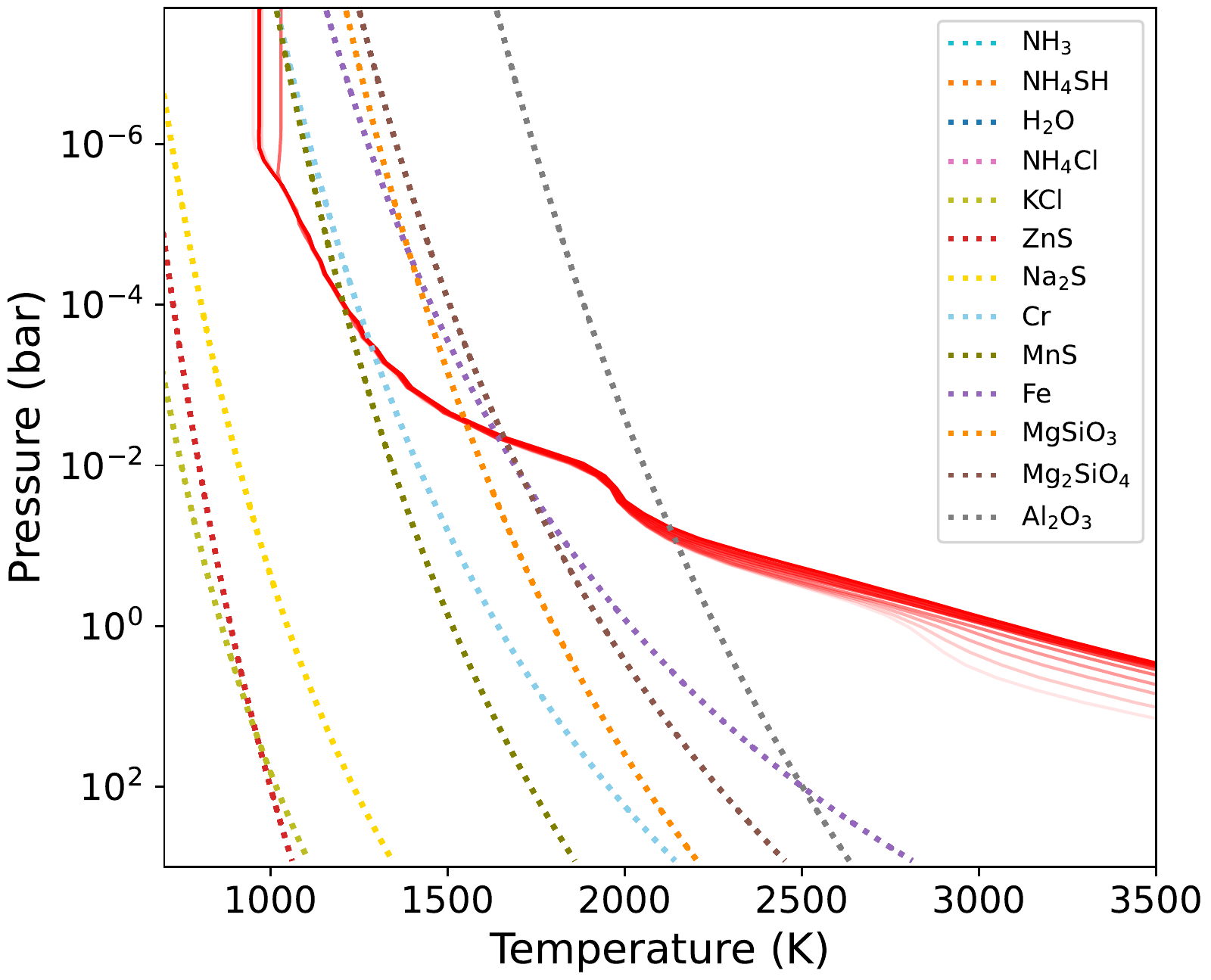}
~
  \includegraphics[width =\columnwidth]{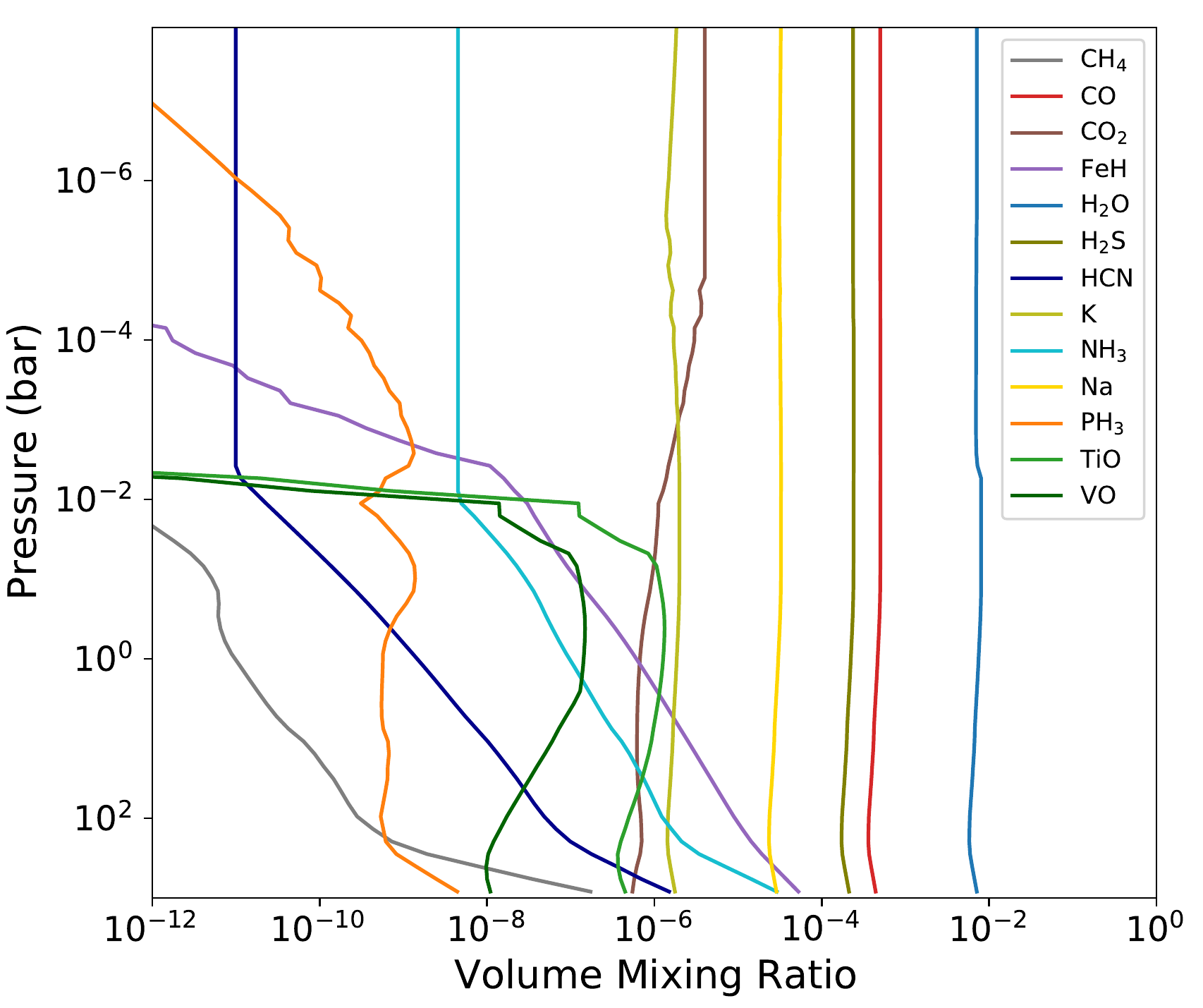}
\caption{Temperature-pressure profiles (solid line) of WASP-79\,b atmosphere (top) assuming radiative transfer equilibrium and condensation curves (dotted lines). The atmospheric structure is computed for different interior temperatures from 350K (light red) to 800K (red). Gas species abundances in the WASP-79\,b atmosphere (bottom) obtained with Exo-REM, assuming a 10 x solar metallicity, non-equilibrium chemistry, an Eddy diffusion coefficient of 10$^{8}$ cm$^2$/s, and an interior temperature of 600K.}
  \label{fig:4ExoREMsimulations}
\end{figure}
Exo-REM is a self-consistent software for brown dwarfs and giant exoplanets atmospheric simulations. 
The stellar and planetary parameters are set to Table \ref{table:1parameters}. 
The light-source spectrum is modelled using PHOENIX \citep{Allard_2012} with an effective temperature of 6600K, a surface gravity of $\log g=4,$ and a solar metallicity. We calculated the structure of the atmosphere of WASP-79\,b using 71~layers between 10$^{-8}$ and 10$^3$ bar. 
We included 13 absorbing species (CH$_4$, CO, CO$_2$, FeH, H$_2$O, H$_2$S, HCN, K, Na, NH$_3$, PH$_3$, TiO, VO) using k-coefficient tables computed with a resolving power of 500 and three collision-induced absorption sources (H$_2$-H$_2$, H$_2$-He, H$_2$O-H$_2$O). The chemistry is allowed to be out of equilibrium for the different species. We used a 10x solar metallicity and a constant Eddy diffusion coefficient of 10$^{8}$ cm$^2$/s. We initialised the temperature-pressure profile to an isothermal temperature profile using the equilibrium temperature of WASP-79 b. Then, we used the results of the first 25 iterations of the retrieval analysis to set the a priori temperature profile. We obtained a solution using a retrieval tolerance for the flux convergence of 0.01.

Figure~\ref{fig:4ExoREMsimulations} shows the calculated abundances in volume-mixing ratios of gas species 
and the temperature pressure profiles of WASP-79\,b atmosphere using Exo-REM between 10$^{-8}$ and 10$^3$ bar 
for various interior temperatures ranging from 350 to 800K. 
The gas abundances are represented for the T-P profile obtained with an interior temperature of 600K.
Moreover, all the simulations are made using a clear atmosphere and a metallicity of ten times the solar value. 
At high altitudes, the main gas species are H$_2$O, CO, and H$_2$S in the atmosphere of WASP-79 b. 
FeH abundance remains below 10$^{-6}$ even at pressure probed by HST WFC3 ($\sim$ 10$^{-1}$ and 10$^{-3}$ bar). 
The temperature profile crosses the condensing lines of Cr, MnS, MgSiO$_3$, and Mg$_2$SiO$_4$ between 10$^{-5}$ and 10$^{-2}$ bar. 
Deeper in the atmosphere, below 10$^{-2}$ bar, different species such as TiO and VO are becoming more abundant with volume mixing ratios reaching 10$^{-6}$ and 10$^{-7}$, respectively. We note that a very hot thermosphere could also explain large transit depths at short wavelengths \citep{Yelle_2004}, and we will consider it as modelling improves in the future.

\begin{figure}[tb]
  \centering
  \includegraphics[width =\columnwidth]{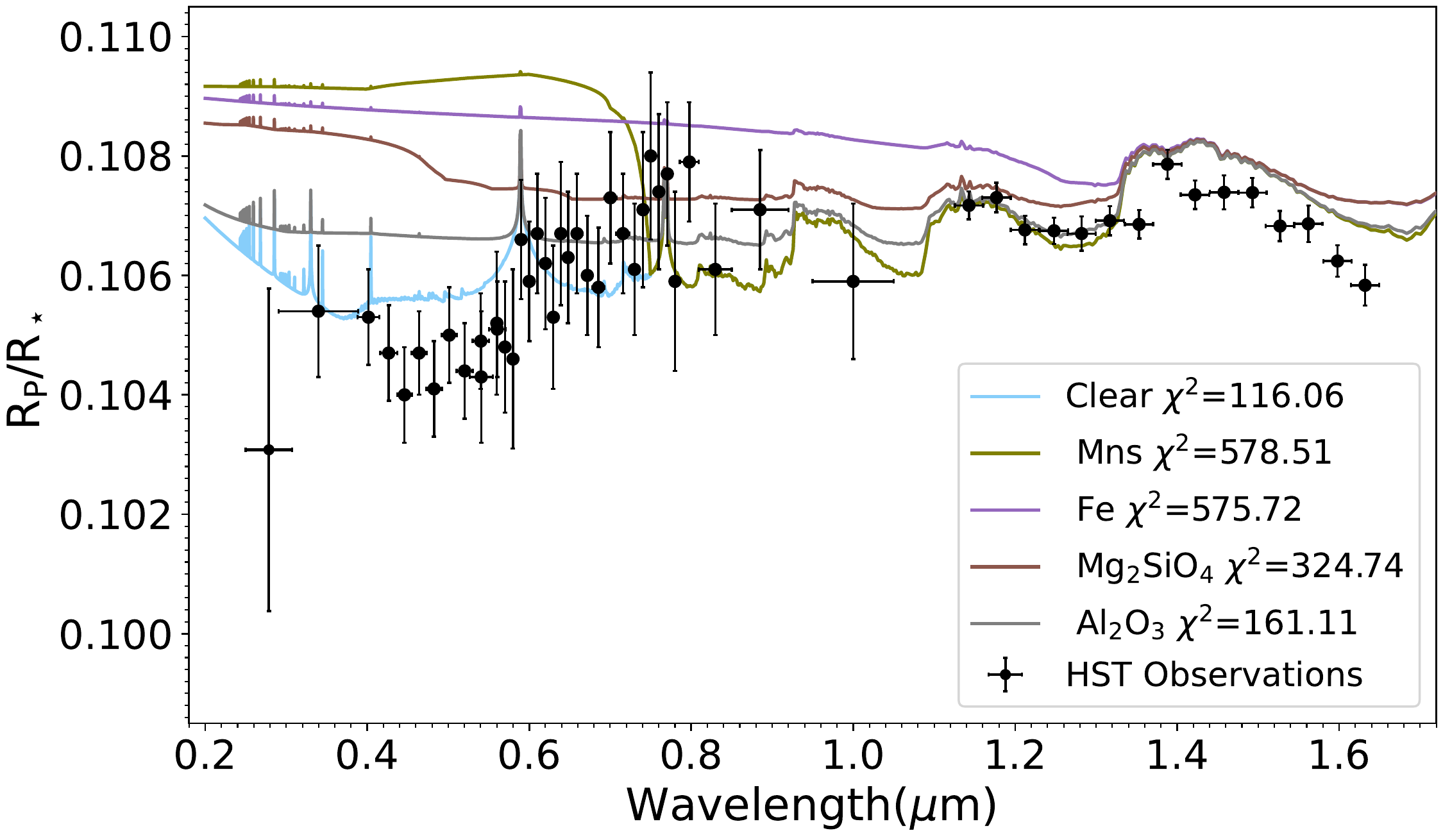}
   \includegraphics[width =\columnwidth]{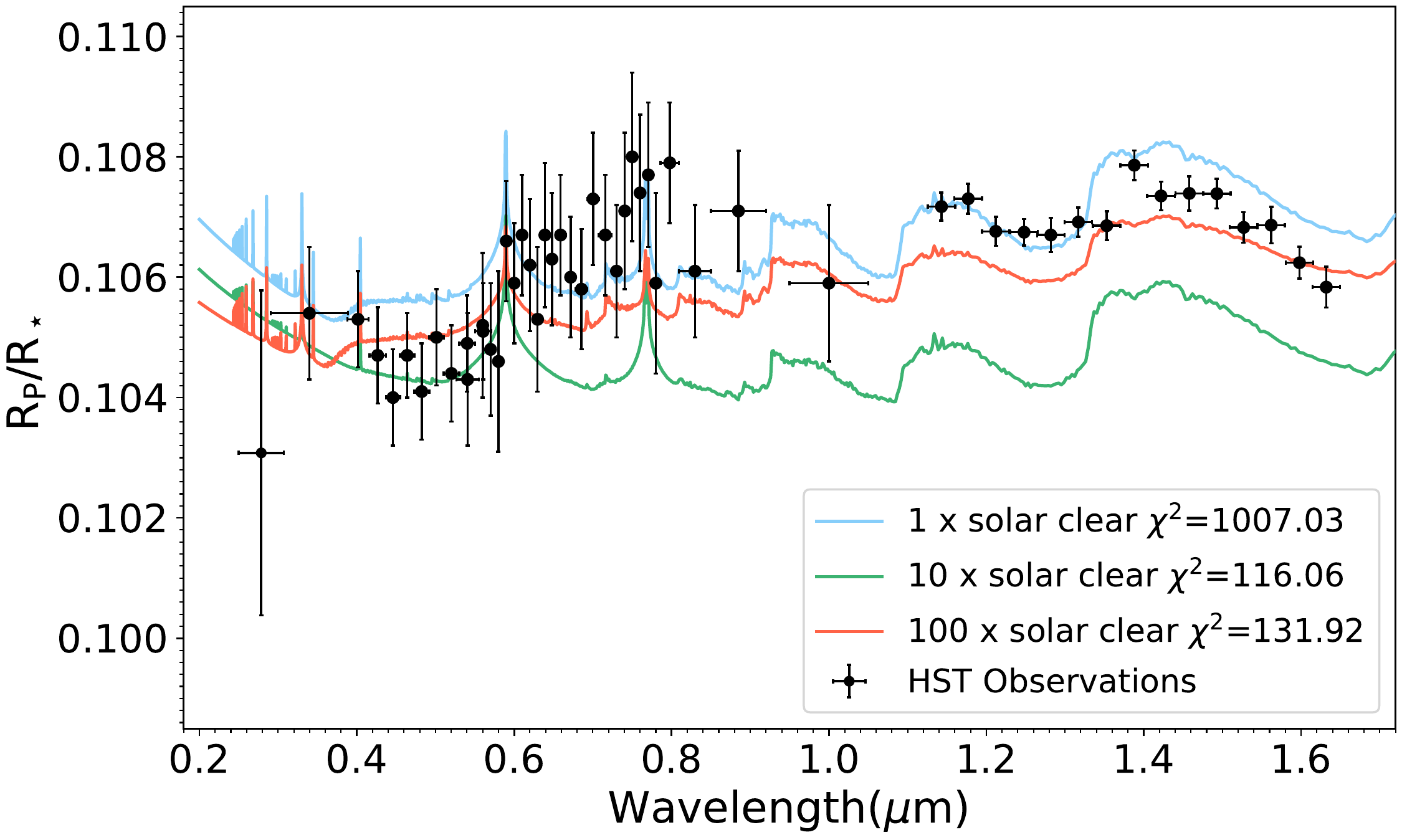}
\caption{WASP-79b HST transmission spectrum observations (black) and simulated spectra using Exo-REM (colours). WASP-79\,b atmosphere is simulated using a 10 x solar metallicity and an interior temperature of 600\,K, and we include different clouds (top). We also tested a clear atmosphere while changing the metallicity of the atmosphere (bottom).}
\label{fig6:exorem_simulations}
\end{figure}

According to the \citet{Gao_2020} study on hot Jupiter cloudiness as a function of temperature, 
the WASP-79\,b spectrum should be dominated by clouds, particularly by Mg$_2$SiO$_4$ around 1700 K. 
After simulating a clear atmosphere with no clouds, we then include different clouds in separate simulations 
with a fixed sedimentation parameter set to 2. 
The metallicity is set to ten times the solar value, and the interior temperature is fixed to 600\,K. 
Fig.~\ref{fig6:exorem_simulations} (top) compares the HST transmission observations with forward models, including different condensing species. We indicate the Chi-squared ($\chi^2$) results for each model. We used our NUV measurements from the broadband analysis on visit \#66 and include the large band value after 2500\AA, similarly to Fig.~\ref{fig:5HST_spectrum}. The value around 2400\AA\  does not appear in Fig.~\ref{fig6:exorem_simulations} for clarity, but is is taken into account for chi-squared computations. 
$\chi^2$ results indicate that forward models tested here fit the HST transmission spectrum of WASP-79 b poorly. However, it is best explained by a clear atmosphere.
We note that none of the clouds presented here can explain the high planet-to-star radius ratio above 0.12 found in both the broad- and narrow-band analyses. 
Fig.~\ref{fig6:exorem_simulations} (bottom) shows the simulated spectra using 1, 10, and 100 times solar metallicity as a comparison. 
The resolution of the spectra in the NUV does not allow us to distinguish between the three different metallicities. 
The observations in the NIR are best explained by the ten times solar metallicity scenario, especially for the absorption feature of water at 1.4$\mu$m.
However, we note that the slope observed after 1.5$\mu$m is not well fitted, and the atmosphere of WASP-79\,b might have a slightly higher metallicity.

\begin{figure}[tb]
  \centering
  \includegraphics[width =\columnwidth]{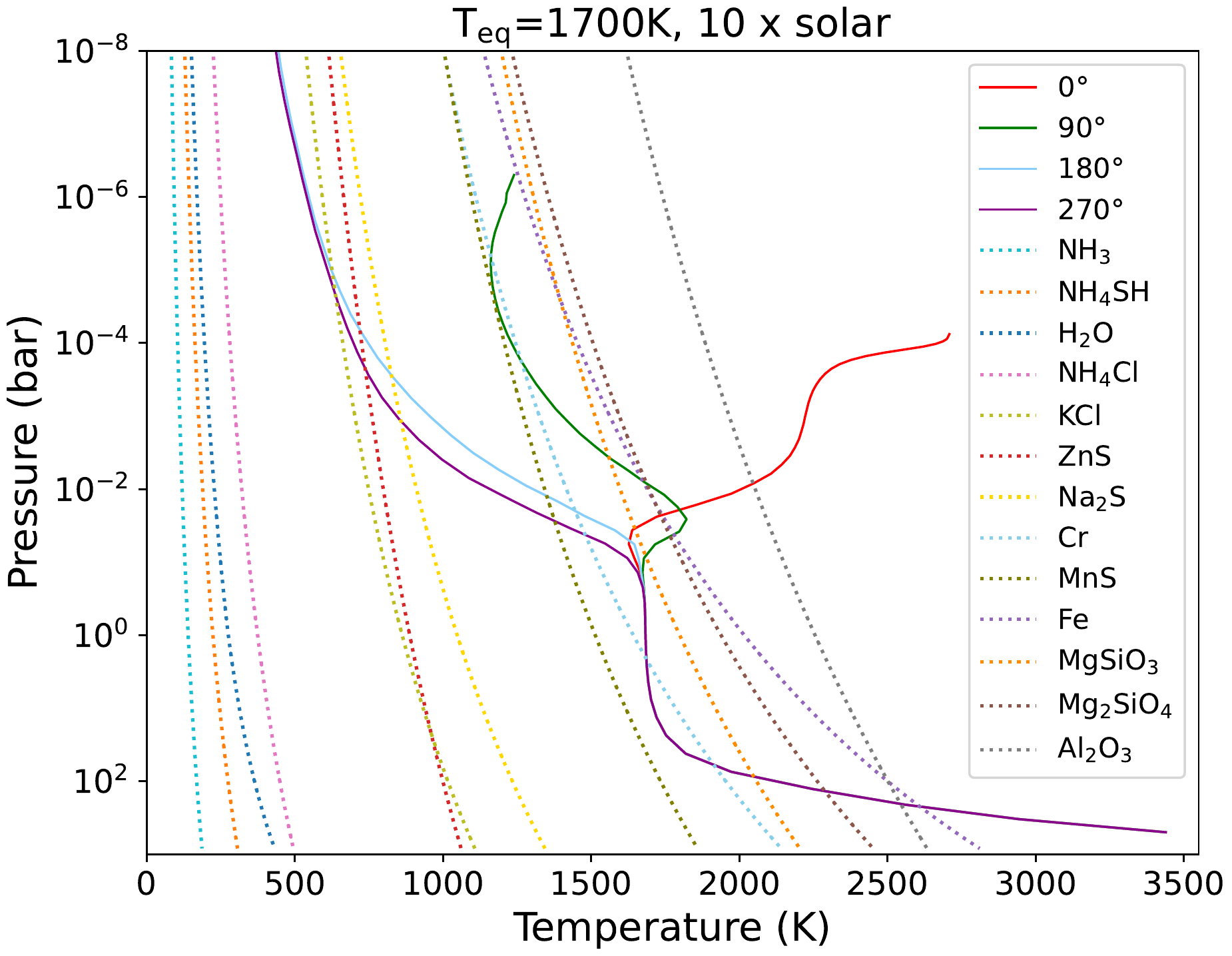}
\caption{Temperature pressure profiles (solid lines) for four longitudes. 
The 0$^{\circ}$ longitude corresponds to the sub-stellar point. The profile is obtained from the grid presented in \citet{Moses_2021} based on the 2D-ATMO circulation model described in \citet{Tremblin_2017} for an effective temperature of 1700 K and a 10x solar metallicity. 
Condensation curves (dotted lines) are from an Exo-REM simulation using the same temperature and metallicity.}
\label{fig7:2d_effects_TP}
\end{figure}

The WASP-79\,b spectrum is consistent with a clear atmosphere, yet the planet could also present a cold, cloudy limb 
and a clear limb on the other side that would differently shape the spectrum. 
We explore this 2D effect using the temperature grid presented in \citet{Moses_2021} based on the 2D-ATMO circulation model described in \citet{Tremblin_2017}. 
Fig.~\ref{fig7:2d_effects_TP} shows the temperature-pressure profiles for four different longitudes, 
where the 0$\degr$ longitude corresponds to the sub-stellar point, using an effective temperature of 1700 K and a metallicity that is ten times the solar value. 
The condensation curves are from Exo-REM simulations of WASP-79 b's atmosphere using the temperature-pressure profiles from the grid of \citet{Moses_2021}. KCl, ZnS, and Na$_2$S could condense on the night's 270$^{\circ}$ limb (purple line). 
As seen before, WASP-79\,b spectrum suggests a clear atmosphere. 
We would then have a clear limb on the day side at 90$^{\circ}$ and a cloudy limb on the other side.

\begin{figure}[htb]
  \centering
  \includegraphics[width =\columnwidth]{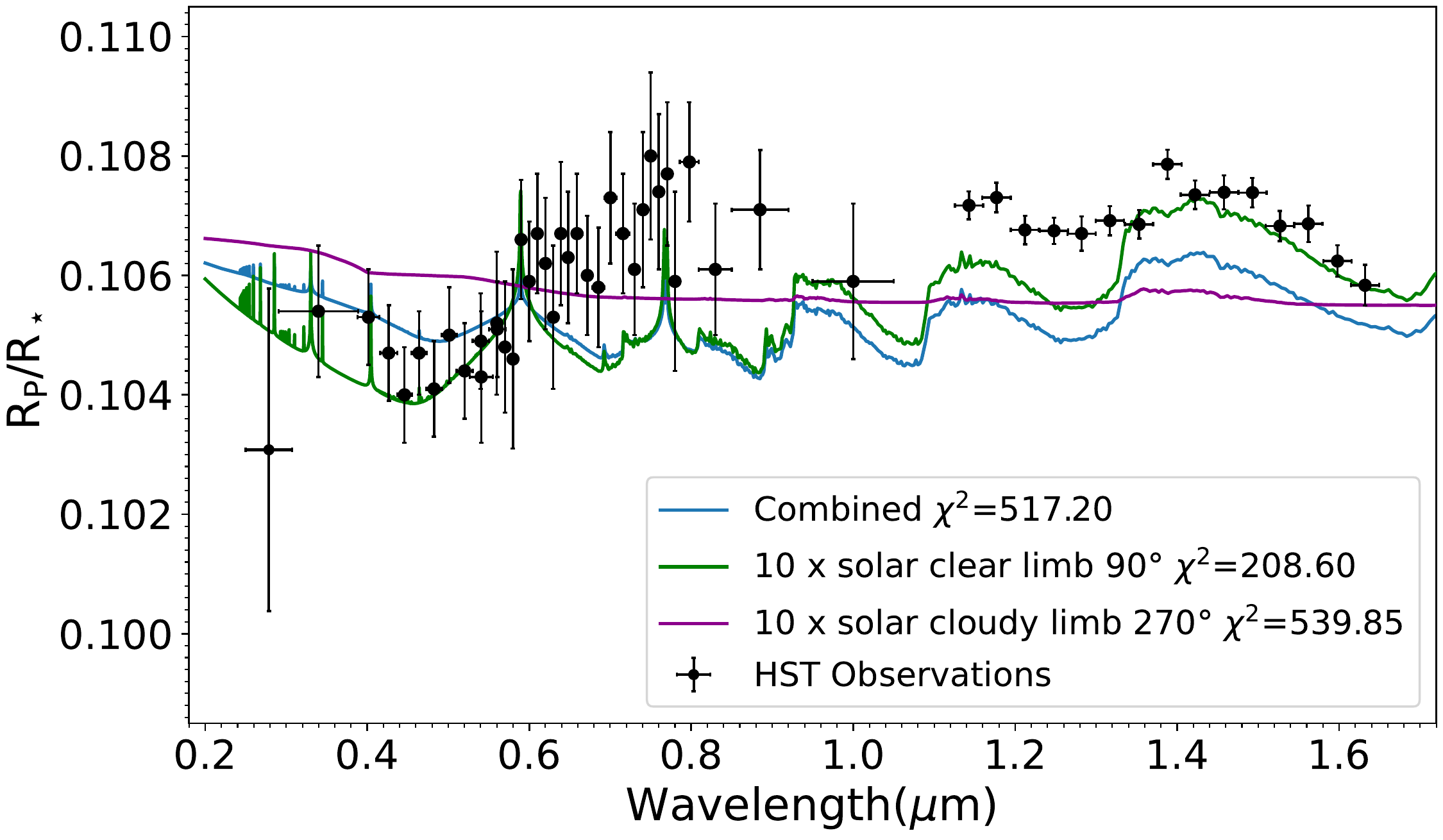}
\caption{WASP-79b HST transmission spectrum observations (black) and simulated spectra using Exo-REM (colours). WASP-79\,b atmosphere is simulated using a 10x solar metallicity and interior temperature of 600 K. We include KCl, ZnS, and Na$_2$S on the night side (purple) simulation and remove clouds on the day side (green). The combined spectrum is in blue.}
\label{fig:A5}
\end{figure}
\begin{figure*}[htpb]
  \centering
  \includegraphics[width =\textwidth]{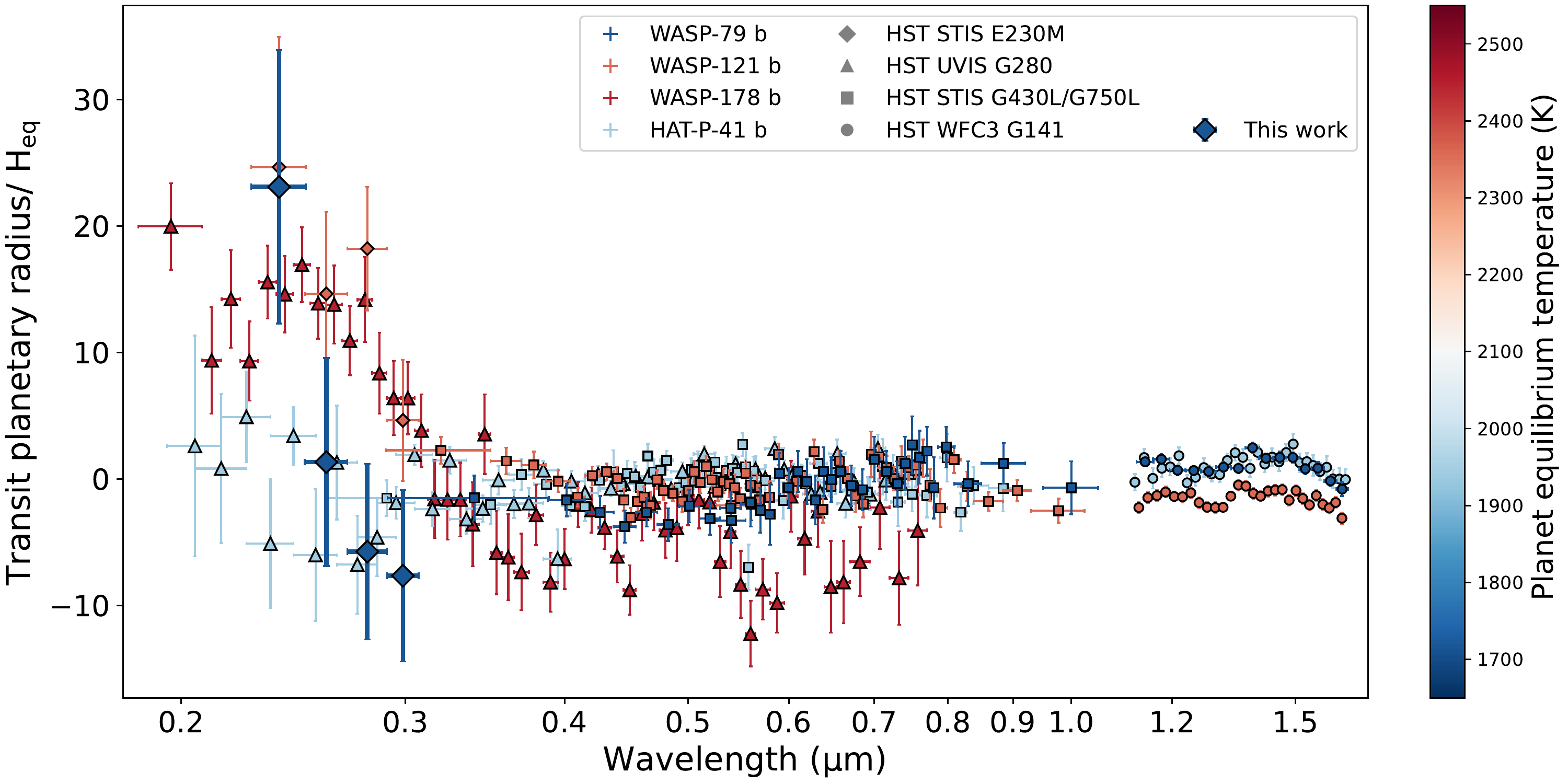}
\caption{NUV-to-NIR transmission spectra normalised by the scale height for WASP-79\,b (T$_{\rm eq}$=1716K) and three additional hot Jupiters: WASP-121\,b (T$_{\rm eq}$=2358K) \citep{Evans_2017, Evans_2018}, HAT-P-41\,b (T$_{\rm eq}$=1941K) \citep{Wakeford_2020, Sheppard_2021}, and WASP-178\,b (T$_{\rm eq}$=2450K) \citep{Lothringer_2022}. WASP-121\,b and WASP-79 b's HST STIS E230M measurements are from this work. The formalism is similar to Figure 2 in \citet{Lothringer_2022}.   }
  \label{fig:HJcomp}
\end{figure*}
We used the two temperature-pressure profiles at longitudes of 90$^{\circ}$ and 270$^{\circ}$ as inputs for WASP-79\,b simulations. 
We included clouds such as KCl, Na$_2$S, and ZnS, for the cloudy limb simulation, and we removed them for the clear limb one. Fig.~\ref{fig:A5} shows the modelled spectra for the clear 
and cloudy cases and the combined spectrum (blue) corresponding to a weighted mean of the two limbs' spectra. 
The combined spectrum does not improve the fit of WASP-79 b; in particular, it does not fit 
the water features around 1.4$\mu$m. However, it must be noted that the 2D grid was built for sub-Neptune planets 
and not for hot Jupiters. 
Besides this, H$^{-}$ was not included in Exo-REM, and therefore it is not present as an opacity source in those simulations,  
although H$^{-}$ was detected in the analyses by \citet{Sotzen_2020} and \citet{rathcke2021hst} of the HST data. 
These species could also be affected by 3D effects, created on one limb and eliminated on the other, causing the spectrum to be shaped differently. 

\subsection{Hot-Jupiter near-UV absorption}
Figure \ref{fig:HJcomp} shows the planetary transit radius normalised by the scale height with respect to the wavelength for four hot Jupiters. We followed the \citet{Lothringer_2022} formalism and added our measurements of the  WASP-121\,b and WASP-79\,b planetary radius using HST STIS E230M (see Figure 2 of \citet{Lothringer_2022}). We used HAT-P-41\,b HST WFC3 UVIS observations from \citet{Wakeford_2020} and HST STIS G430L, G750L and HST WFC3 G141 from \citet{Sheppard_2021}. 
WASP-121\,b's HST STIS G430L, G750L and WFC3 G141 observations are from \citet{Evans_2017,Evans_2018}, while WASP-178\,b WFC3 UVIS G280 measurements are from \citet{Lothringer_2022}. Surprisingly, WASP-79 b's transit spectrum appears to show an absorption in the NUV similar to WASP-121\,b and WASP-178\,b,  while their equilibrium temperatures are higher ($\sim$2350\,K and $\sim$2450\,K for 
WASP-121\,b and WASP-178\,b, respectively). 
Among those, WASP-121\,b's has the largest increase in the NUV absorption depth, with a depth increase more than 30~times the atmospheric scale height. 
Conversely, the transit spectrum of HAT-P-41 b shows no increase in the absorption depth in the NUV despite its intermediate temperature (1950 K).

\citet{Lothringer_2022} interpreted the increase in the planet-to-star radius ratio at short wavelengths by the absorption of SiO, a precursor of condensate clouds, using simulations based on equilibrium chemistry. Comparing WASP-121 b, HAT-P-41\,b and WASP-178\,b transmission spectra put a first temperature constraint on silicate cloud formation. The condensation could begin on exoplanets with effective temperatures between 1950 and 2450K. At short wavelengths, WASP-79 b displays a similar feature to WASP-121 b and WASP-78 b. Our large absorption measurement in the atmosphere of WASP-79 b, with an equilibrium temperature of 1716 K, challenges this direct interpretation. We showed that none of the strong absorption lines in the narrow-band analysis were attributed to iron. Thus, the large absorption feature below 2400\,\AA\ could be interpreted as SiO absorption. 
%Theoretical models at equilibrium support this theory. 
The extended Figure 3 in \citet{Lothringer_2022} shows the partial pressures of iron- and silicon-bearing species as a function of temperature. Silicates and iron condense between 1500 and 2000K, between 1\,mbar and 10 bar for atmospheres with metallicities between one and ten times solar. The lack of absorption in the atmosphere of HAT-P-41\,b was explained in \citet{Lothringer_2022} by the rainout of refractory species from the gas phase. We observed an absorption for WASP-79 b as high as for WASP-121\,b and WASP-178\,b (see Fig.~\ref{fig:HJcomp}), making it one of the most important spectral features observed in an exoplanet in terms of atmospheric scale heights at this temperature. How WASP-79 b avoided rainout whereas HAT-P-41 b did not remains puzzling.

\section{Conclusion}
We obtained a NUV transmission spectrum of WASP-79\,b's atmosphere using new HST STIS E230M data. 
We found an increase in the transit depth at short wavelengths (below $\la$2600\AA ). 
A narrow-band transmission spectrum at a resolution of 4~\text{\AA} did not reveal particular absorption lines but confirmed the global increase in the planet's apparent radius at short wavelengths. 
Contrary to the exoplanet WASP-121~b, the highest values of the planet radius in the narrow-band transmission spectrum do not exceed the equivalent radius 
of the Roche lobe, but they reache about 75$\%$ of its value. 
The absorption observed below 2500\AA\ corresponds to about 44\% of the Roche-lobe equivalent radius.  

A rapid and straightforward evaluation of the possible impact of spot and facul\ae\ on the transmission spectrum is performed in the discussion section using a blackbody spectrum. A more realistic evaluation of the fraction of the stellar area covered by the spots or facul\ae\ could be assessed using an atmospheric model. However, non-solar-type stars' spectra and facul\ae\ are poorly known. Our first approximation shows that the effect of stellar activity on the planet's radius is negligible and does not explain the amplitude of the observed features. Given the order of magnitude, a more advanced model would not affect the overall result, and the conclusion would remain similar.  

A 1D simulation of the deeper layers of the atmosphere was performed using Exo-REM 
with non-equilibrium chemistry and a ten-times-solar metallicity. The temperature pressure profile crosses condensation curves of radiatively active clouds such as MnS, Fe, Mg$_2$SiO$_4$, or Al$_2$O$_3$. 
Yet, none of those absorbing species can explain the observed increase in the planet's radius at short wavelengths. 
The overall HST transmission spectrum suggests a clear atmosphere for WASP-79\,b, but the planet might be tidally locked, and 3D effects could play an important role. We explored the 2D effects using the temperature-pressure profiles grid 
from \citet{Moses_2021} based on 2D-ATMO of \citet{Tremblin_2017}. Clouds made of KCl, Na$_2$S, and ZnS could be created on one side and evaporated on the other. 

The comparison of WASP-79 b's transmission spectrum with three other hot Jupiters at short wavelengths shows a surprisingly similar absorption around 2400\,\AA . 
While the HAT-P-41\,b spectrum is flat, WASP-79\,b, WASP-121\,b, and WASP-178\,b display large absorption features between 0.2 and 0.3$\mu$m. 
This has been interpreted as SiO absorption by \citet{Lothringer_2022} in the atmosphere of WASP-178\,b. 
WASP-79\,b's NUV excess absorption corresponds to a scale height of more than 20, making it one of the largest spectral features observed in an exoplanet at this temperature (1716K). 
If this feature is attributed to absorption by SiO, silicate cloud formation must be investigated to understand the disparity in this sample of hot Jupiters.

Further observations of WASP-79\,b in the UV could better characterise this absorption. A combination of the present STIS E230M measurement with UVIS observation could be of great interest, as suggested in \citet{Lothringer_2022}. Ground-based high-resolution observations of WASP-79 b 's atmosphere could also detect species such as Fe or atomic Si and help us understand whether species are raining out. An increase in the number of exoplanets observed in the UV will help to investigate the cloud formation in hot Jupiters.

\begin{acknowledgements}
This work is based on observations made with the NASA/ESA Hubble Space Telescope obtained at the Space Telescope.
Science Institute, which is operated by the Association of Universities for Research in Astronomy, Inc. 
Support for this work was provided by NASA through grants under the HST-GO-14767 program from the STScI.
A.G. and A.L.d.E. acknowledge support from the CNES (Centre national d'\'etudes spatiales, France). This work has been carried out in the frame of the National Centre for Competence in Research PlanetS supported by the Swiss National Science Foundation (SNSF). This project has received funding from the European Research Council (ERC) under the European Union's Horizon 2020 research and innovation programme (project {\sc Spice Dune}, grant agreement No 947634). 

G.W.H. acknowledges long-term support of the APT program from NASA, NSF, Tennessee State University, and the State of
Tennessee through its Centers of Excellence Program.

J.K.B is supported by an STFC Ernest Rutherford Fellowship (ST/T004479/1).

The authors thank Hannah Wakeford for constructive remarks that improved our study.
\end{acknowledgements}

\bibliographystyle{aa}
\bibliography{main.bib}

\end{document}